\documentclass[dvipsnames]{elsarticle}
\usepackage{siunitx}
\usepackage{hyperref}
\usepackage{graphicx}  
\usepackage{dcolumn}   
\usepackage{bm}        
\usepackage{amssymb}   
\usepackage{amsfonts}
\usepackage{graphics}
\usepackage{grffile}   
\usepackage{epsfig}
\usepackage[usenames]{color}
\usepackage[normalem]{ulem} 
\usepackage{color}
\usepackage[utf8]{inputenc}
\usepackage[T1]{fontenc}
\usepackage{amsmath}
\usepackage{enumitem}

\usepackage{subcaption}
\usepackage{adjustbox}

\usepackage{lineno}

\DeclareSIUnit\ele{e^{\text{-}}}
\DeclareUnicodeCharacter{2212}{-}

\usepackage{lineno}

\usepackage{xcolor}
\usepackage{tikz}
\usepackage{environ}
\makeatletter
    \newsavebox{\measure@tikzpicture}
        \NewEnviron{scaletikzpicturetowidth}[1]{%
            \def\tikz@width{#1}%
            \def\tikzscale{1}\begin{lrbox}{\measure@tikzpicture}%
            \BODY
            \end{lrbox}%
            \pgfmathparse{#1/\wd\measure@tikzpicture}%
            \edef\tikzscale{\pgfmathresult}%
            \BODY
        }
\makeatother
\usetikzlibrary{math}
\usetikzlibrary{calc}
\usetikzlibrary{decorations.pathreplacing, decorations.markings}
\usetikzlibrary{patterns, arrows, arrows.meta}

\usepackage{imakeidx}
\makeindex[columns=3, title=Alphabetical Index, intoc]
\makeindex

\journal{Nucl. Instrum. Methods Phys. Res. A}

\begin{document}
\begin{frontmatter}
\title{Time performance of Analog Pixel Test Structures with in-chip operational amplifier implemented in 65~nm CMOS imaging process}


\author[CERN]{Gianluca Aglieri Rinella}
\author[unito,infnTO]{Luca Aglietta}
\author[infnTS]{Matias Antonelli}
\author[uniba,infnBA]{Francesco Barile}
\author[infnTO]{Franco Benotto}
\author[unito,infnTO]{Stefania Maria Beolè}
\author[unito,infnTO]{Elena Botta}
\author[poliBA,infnBA]{Giuseppe Eugenio Bruno}
\author[CERN]{Francesca Carnesecchi}
\author[uniba,infnBA]{Domenico Colella}
\author[uniba,infnBA]{Angelo Colelli}
\author[units,infnTS]{Giacomo Contin}
\author[infnBA]{Giuseppe De Robertis}
\author[infnTO]{Floarea Dumitrache}
\author[infnBA]{Domenico Elia}
\author[polito,infnTO]{Chiara Ferrero}
\author[nikhef]{Martin Fransen}
\author[CERN]{Alex Kluge}
\author[infnBA]{Shyam Kumar}
\author[CERN,IPHC]{Corentin Lemoine}
\author[infnBA]{Francesco Licciulli}
\author[unito,infnTO]{Bong-Hwi Lim}
\author[infnBA]{Flavio Loddo}
\author[CERN]{Magnus Mager}
\author[unica,infnCA]{Davide Marras}
\author[CERN]{Paolo Martinengo}
\author[infnBA]{Cosimo Pastore}
\author[infnBA,uniJAM]{Rajendra Nath Patra}
\author[unito,infnTO]{Stefania Perciballi}
\author[CERN]{Francesco Piro}
\author[infnTO]{Francesco Prino}
\author[infnTO,UPO]{Luciano Ramello}
\author[infnBA]{Arianna Grisel Torres Ramos}
\author[CERN]{Felix Reidt}
\author[nikhef]{Roberto Russo}
\author[CERN]{Valerio Sarritzu}
\author[unito,infnTO]{Umberto Savino\corref{cor1}}
\author[heidel]{David Schledewitz}
\author[nikhef]{Mariia Selina}
\author[IPHC]{Serhiy Senyukov}
\author[infnTO,UPO]{Mario Sitta}
\author[CERN]{Walter Snoeys}
\author[nikhef]{Jory Sonneveld}
\author[CERN]{Miljenko Suljic}
\author[poliBA,infnBA]{Triloki Triloki}
\author[unito,infnTO]{Andrea Turcato}

\affiliation[unito]{organization={University of Torino - Department of Physics},
city={Torino},
country={Italy}
}
\affiliation[infnTO]{organization={INFN, Sezione di Torino},
city={Torino},
country={Italy}
}
\affiliation[polito]{organization={Polytechnic of Torino - DET Department},
city={Torino},
country={Italy}
}
\affiliation[unica]{organization={University of Cagliari},
city={Cagliari},
country={Italy}
}
\affiliation[infnCA]{organization={INFN, Sezione di Cagliari},
city={Cagliari},
country={Italy}
}
\affiliation[units]{organization={University of Trieste},
city={Trieste},
country={Italy}
}
\affiliation[infnTS]{organization={INFN, Sezione di Trieste},
city={Trieste},
country={Italy}
}

\affiliation[CERN]{organization={European Organisation for Nuclear Research (CERN)},
city={Geneva},
country={Switzerland}
}
\affiliation[uniba]{organization={University of Bari - Department of Physics},
city={Bari},
country={Italy}
}
\affiliation[infnBA]{organization={INFN, Sezione di Bari},
city={Bari},
country={Italy}
}
\affiliation[poliBA]{organization={Polytechnic of Bari - Department of Physics DIF}, 
city={Bari},
country={Italy}
}
\affiliation[IPHC]{organization={Université de Strasbourg, CNRS, IPHC UMR 7178}, 
city={Strasbourg}, 
country={France}
}
\affiliation[nikhef]{organization={National Institute for Subatomic Physics (Nikhef)},
city={Amsterdam},
country={Netherlands}
}
\affiliation[UPO]{organization={Università del Piemonte
Orientale},
city={Vercelli},
country={Italy}
}
\affiliation[uniJAM]{organization={Department of Physics, University of Jammu}, 
city={Jammu}, 
country={India}
}
\affiliation[heidel]{organization={Ruprecht Karls Universität Heidelberg},
city={Heidelberg},
country={Germany}
}

\cortext[cor1]{Corresponding author: umberto.savino@unito.it}

\begin{abstract}
In the context of the CERN EP R\&D on monolithic sensors and the ALICE ITS3 upgrade, the Tower Partners Semiconductor Co (TPSCo) \SI{65}{nm} process has been qualified for use in high energy physics, and adopted for the ALICE ITS3 upgrade. 
An Analog Pixel Test Structure (APTS) featuring fast per pixel operational-amplifier-based buffering for a small matrix of four by four pixels, with a sensor with a small collection electrode and a very non-uniform electric field, was designed to allow detailed characterization of the pixel performance in this technology. 
Several variants of this chip with different pixel designs 
have been characterized with a \(120 \text{ GeV/\it{c}}\) positive hadron beam.
This result 
indicates that the APTS-OA prototype variants with the best performance achieve a time resolution of \SI{63}{ps} with a detection efficiency exceeding 99\% and a spatial resolution of \SI{2}{\um}, highlighting the potential of TPSCo 65nm CMOS imaging technology for high-energy physics and other fields requiring precise time measurement, high detection efficiency, and excellent spatial resolution.



\end{abstract}


\end{frontmatter}

%
%
\tikzset{
	beamarrow/.style={
		decoration={
			markings,mark=at position 1 with 
			{\arrow[scale=2,>=stealth]{>}}
		},postaction={decorate}
	}
}
\tikzset{
	pics/.cd,
	vector out/.style={
		code={
		\draw[#1, thick] (0,0)  circle (0.15) (45:0.15) -- (225:0.15) (135:0.15) -- (315:0.15);
		}
	}
}
\tikzset{
	pics/.cd,
	vector in/.style={
		code={
		\draw[#1, thick] (0,0)  circle (0.15);
		 \fill[#1] (0,0)  circle (.05);
		 }
	}
}
\tikzset{
	global scale/.style={
		scale=#1,
		every node/.style={scale=#1}
	}
}
\def\centerarc[#1] (#2)(#3:#4:#5) 
	 { \draw[#1] ($(#2)+({#5*cos(#3)},{#5*sin(#3)})$) arc (#3:#4:#5); }
\section{Introduction}
\label{sec:intro}
In the field of high-energy physics, Monolithic Active Pixel Sensors (MAPS) have emerged as a promising technology for vertex detectors, demonstrating potential to revolutionize tracking capabilities in future experiments~\cite{CONTIN201860, REIDT2022166632}. Their innovative design, integrating readout circuitry with the detection volume in single, commercially fabricated chips, facilitates ultra-thin, large-scale tracking detectors – a breakthrough for precision and scalability in research.

As high-energy physics experiments evolve, they are expected to cope with higher interaction rates and 
to resolve several primary interaction vertices as well as vertices from secondary decays~\cite{Aberle:2749422}.  
This trend demands pixel detectors with enhanced spatial resolution and superior timing capabilities, essential for precise tracking in dense particle environments~\cite{AglieriRinella:2808204}. 
Recent developments in the 180 nm TowerJazz CMOS technology have led to significant improvements. 
Sensor modifications have enabled improved depletion~\cite{SNOEYS201790} and significantly accelerated charge collection especially for hits near the pixel edge~\cite{Munker_2019}, allowing for time resolutions between 100 and 120 ps~\cite{KUGATHASAN2020164461, Braach_2023}.

A new generation of MAPS detectors is being developed for the third generation ALICE Inner Tracking System (ITS3)~\cite{eprndMAPS,loi,doi:10.7566/JPSCP.34.010011}, using the Tower Partners Semiconductor Co. (TPSCo) \SI{65}{nm} CMOS Image Sensor Process~\cite{towerjazz} introducing similar sensor modifications as in the \SI{180}{nm} technology~\cite{Munker_2019,Braach_2023}. 
This \SI{65}{nm} technology enables higher circuit densities and hence more advanced readout circuitry can be fitted in the same pixel pitch, allowing for more precise timing measurements.
An Analog Pixel Test Structure (APTS), equipped with fast individual Operational Amplifier (OA)-based buffering was produced to evaluate the intrinsic timing performance of the sensors in this new technology.
Several variants of the chip with different pixel designs were implemented and characterized in Ref.~\cite{APTSSF_2023,SnoeysPixel2022} 
and are summarized in the next section.
The results of a study
carried out during a test beam campaign with positive hadrons are presented in this paper, with a focus on the sensor timing performance.
\section{The APTS-OA}
\label{sec:apts_chip}

\begin{figure}[!ht]
    \begin{minipage}[b]{.45\linewidth}
        \centering
        \includegraphics[width=\textwidth]{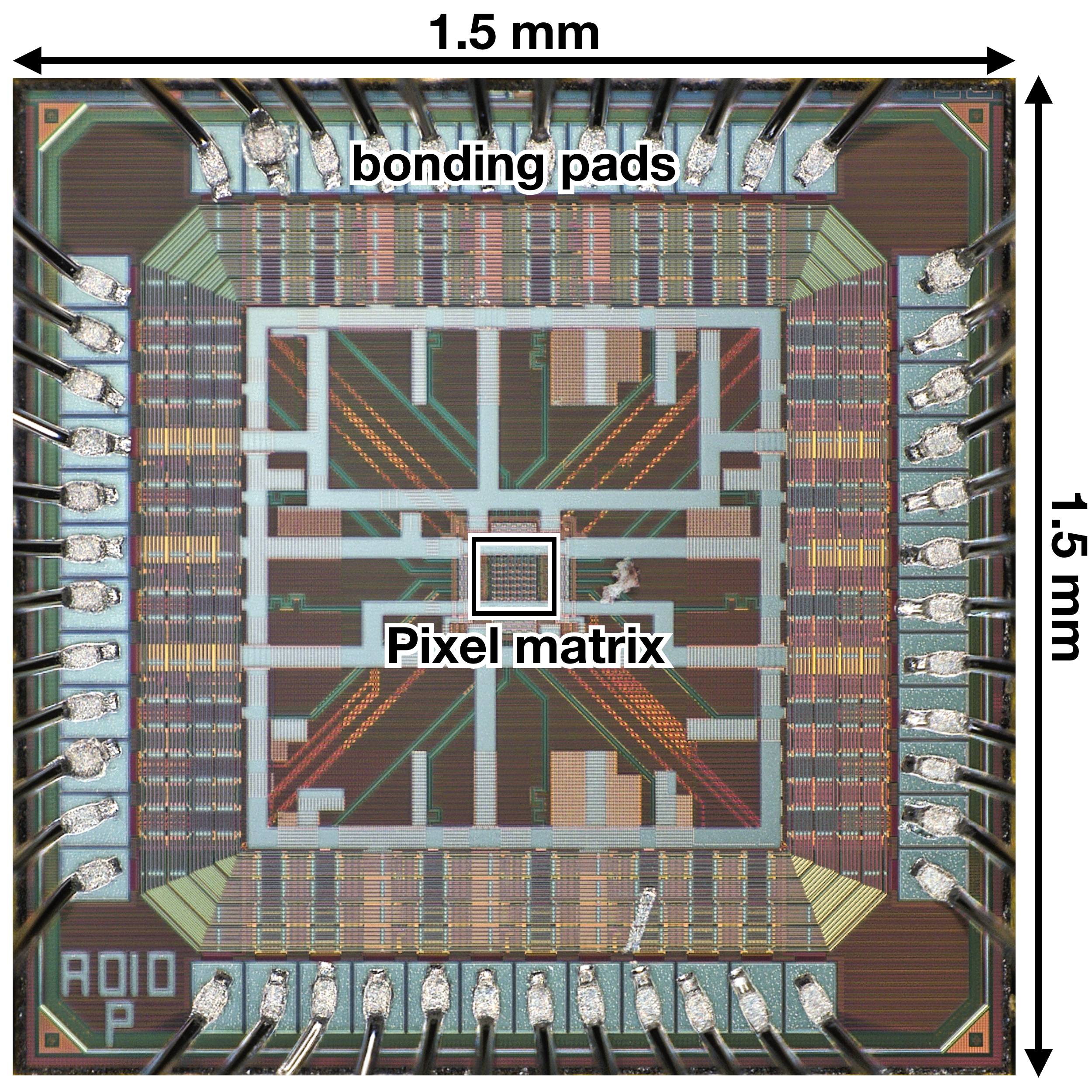}
        \caption{Photograph of the APTS-OA taken with a microscope.}
        \label{fig:APTSphoto}
    \end{minipage}
    \begin{minipage}[b]{.54\linewidth}
        \begin{adjustbox}{width=\textwidth,keepaspectratio}
            \hspace{0.2cm}
            \begin{tabular}{@{}ll@{}}
            \hline
            \multicolumn{2}{c}{APTS-OA}                      \\ \hline
            Die size & $1.5$ mm $\times$ $1.5$ mm                       \\
            Matrix   & $4 \times 4$                                 \\
            Readout  & Direct analogue of central $4\times4$ pixels \\
            Pitch    & $10$~ \textmu m                                 \\
            Process   & \begin{minipage}[t]{0.7\textwidth}
                \begin{itemize}[leftmargin=0pt,itemsep=0pt]
                \item[] Standard
                \item[] Modified
                \item[] Modified with gap
            \end{itemize}
            \end{minipage}
               
            \\ \hline
            \end{tabular}
        \end{adjustbox}
    \captionof{table}{Main characteristics of the APTS-OA.}
    \label{tab:APTSOA}
    \end{minipage}
\end{figure}

The APTS-OA measures \(1.5 \times 1.5 \text{ mm}^2\) and features a \(4 \times 4\) squared \SI{10}{\um} pitch pixel matrix directly buffered to output pads for readout and controlled by a set of external reference currents and voltages. 
This matrix is surrounded by dummy pixels that help to reduce the electric field distortion effects at the edges of the matrix.
In Figure~\ref{fig:APTSphoto}, the APTS-OA chip is shown where, going from the center to the periphery, the pixel matrix, and the bonding pads can be observed. Table~\ref{tab:APTSOA} presents the primary characteristics of the chip detailed below.

The sensor can be reverse biased by applying a negative voltage referred to as \textit{$V_{sub}$} to the substrate and p-wells 
down to a minimum of \SI{-4.8}{V}.

\subsection{Sensor variants}

\begin{figure}[!ht]
\centering
\begin{subfigure}{.9\textwidth}
  \centering
  \includegraphics[width=\textwidth]{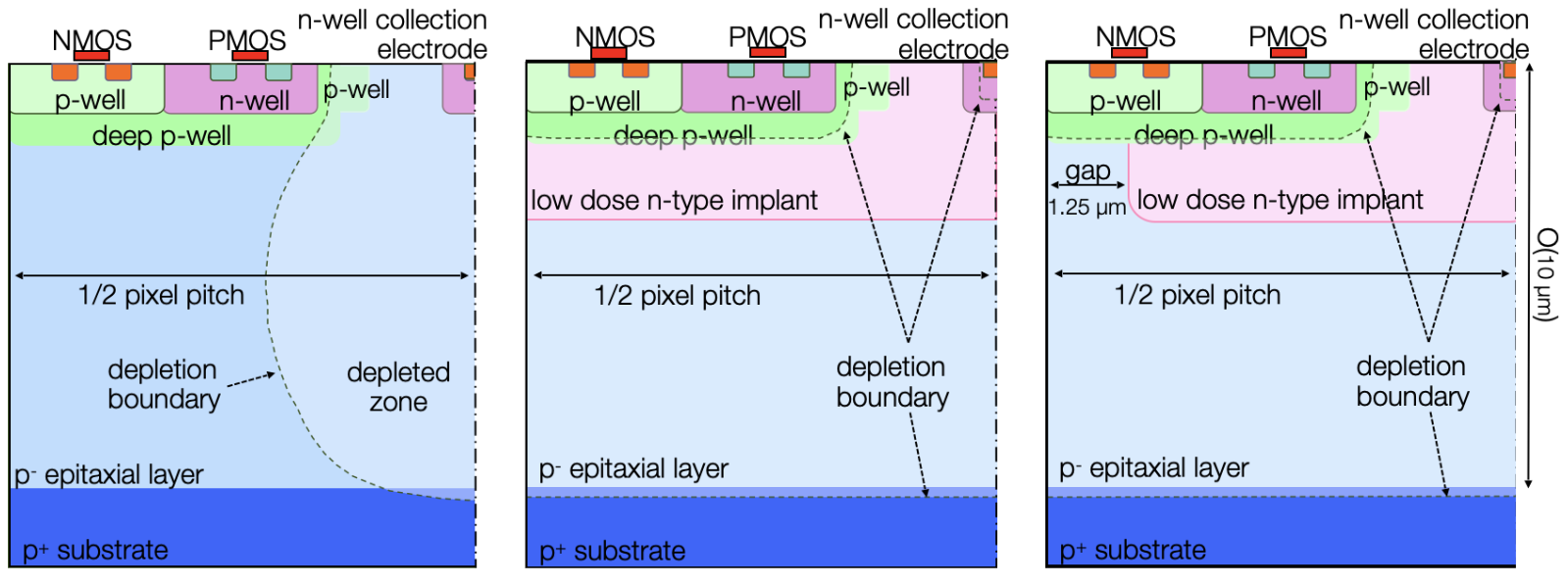}
  \label{fig:sfig1c}
\end{subfigure}
  \captionsetup{justification=centering}
  \caption{Schematic cross section of the three pixel process variants of the APTS. From left to right: standard, modified featuring a low dose deep n-type implant, and modified with gap employing a low dose deep n-type implant with a gap at the pixel's edges. Not to scale.~\cite{APTSSF_2023}}
  \label{fig:crossect}
\end{figure}

The APTS was produced with three process variants of its sensor, starting from a standard imaging CMOS design and introducing modifications resulting in faster charge collection and reduced charge sharing~\cite{AGLIERIRINELLA2023168589,APTSSF_2023,SNOEYS201790}.
\begin{itemize}
\item \textbf{The standard process} (cf. Fig.~\ref{fig:crossect} left) is formed by an n-well collection electrode of about \SI{1}{\um} diameter on a high-resistivity p-type epitaxial layer, grown on top of a low-resistivity p-type substrate. In this structure, the depletion extends from the small n-well electrode resulting in a balloon-shaped depleted zone covering only part of the epitaxial layer. Charge collection outside this region happens mostly by diffusion, leading to long and non-uniform collection times but also to significant charge sharing \cite{MunkerPhD}. 
A deep p-well, located at about \SI{2}{\um} from the collection electrode, separates circuitry n-wells from the epitaxial layer, allowing to use PMOSs as well as NMOSs in the pixel matrix without competing with the electrode for charge collection.
\item \textbf{The modified with blanket implant process}, for brevity called ‘modified’ in the following, (cf. Fig.~\ref{fig:crossect} center) differs from the standard structure for the addition of a low-dose n-type implant in the epitaxial layer, resulting in a planar junction over the full pixel that helps in achieving full depletion of the epitaxial layer. 
\item \textbf{The modified with an implant with gap process}, called ‘modified with gap’ in the following, (cf. Fig.~\ref{fig:crossect} right) is similar to the modified process but the deep n-type implant no longer covers the full pixel: a gap is introduced near the pixel edge to create a lateral field pushing the signal charge towards the collection electrode,
thus decreasing collection time 
(as demonstrated in the simulation~\cite{Munker_2019} for the 180 nm CMOS TowerJazz technology) and charge sharing. 
\end{itemize}
Characterization of a version of the APTS with source follower output buffers (APTS-SF) has demonstrated the improvements described above for the \SI{65}{nm} technology~\cite{Deng_2023,APTSSF_2023}.

\subsection{Analog circuit}

\begin{figure}[!ht]
    \centering
    \includegraphics[width=\textwidth]{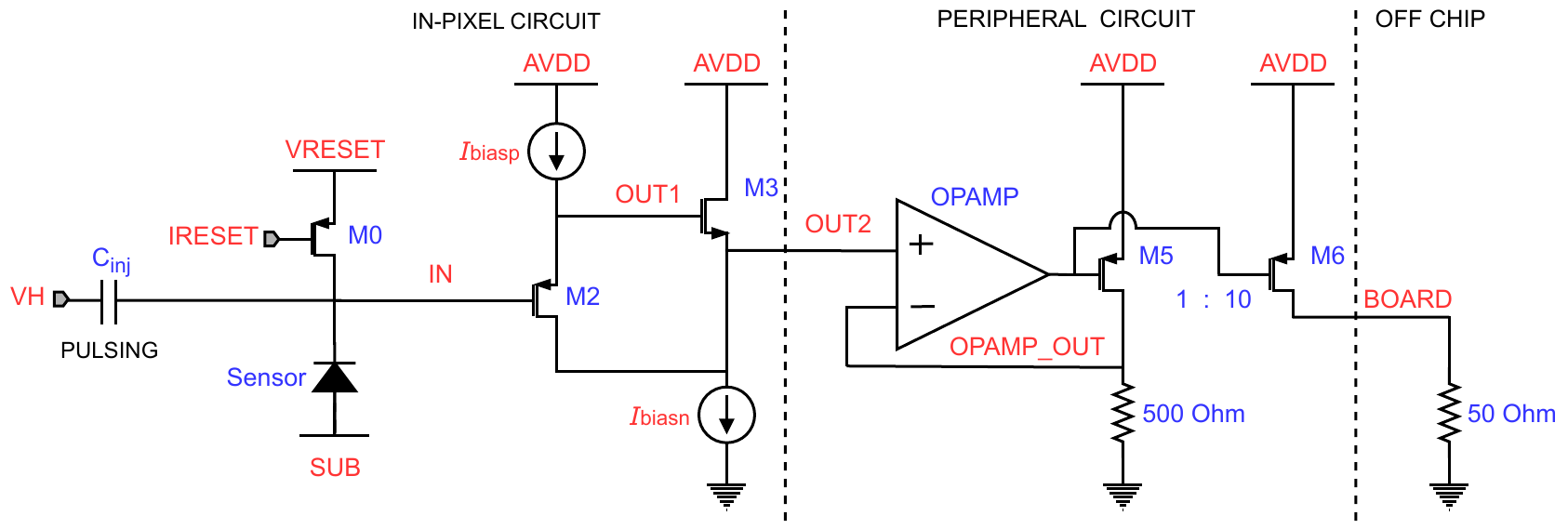}
    \caption{
    Left: APTS-OA front-end circuit with the circuitry inside the pixel. Right: circuitry located in the periphery of the matrix. The off chip line to the measurement system is also represented.}
    \label{fig:frontend}
\end{figure}

The analog circuit of APTS-OA is depicted in Fig.~\ref{fig:frontend}.
An analog front-end powered at 1.2 V ($AVDD$) is connected to the collection electrode of each pixel ($IN$) and buffers its signal to an output pad ($BOARD$). To do so, the circuit is composed of several buffering stages: a low input capacitance stage with two source followers buffering the $IN$ signal to $OUT1$ and $OUT2$, respectively, and a fast operational amplifier driving the output stage $BOARD$. The circuit also provides reset current to the sensor and allows to inject charge at the input of the front-end via a pulsing circuitry. Most of the analog circuit is implemented in the pixel matrix, but due to lack of area the operational amplifiers and output stages are located at the periphery immediately surrounding the pixel matrix.

In case there is no hit, the transistor $M0$ only compensates for the leakage current of the sensor and maintains its collection electrode at a voltage close to $V_{\textrm{reset}}$. 
The current generated after a particle hit entails a voltage drop on the input node ($IN$), turning on the transistor $M0$ to deliver constant current $I_{\textrm{reset}}$.
$I_{\textrm{reset}}$ is chosen larger than the sensor leakage current to ensure that $IN$ returns to its baseline within a few microseconds.

The collection electrode ($IN$) is directly connected to the input of a PMOS source follower stage $M2$ that makes its output $OUT1$ follow the input voltage. 
To take full advantage of the low sensor capacitance in this process, the input capacitance of the front-end is further reduced by making the drain of $M2$ also follow the same voltage.
This is achieved by the NMOS source follower $M3$ driving $OUT2$. These source followers are biased by two current sources: $I_{\textrm{biasp}}$ and $I_{\textrm{biasn}}$.

An in-pixel pulsing circuit can inject charge in the collection electrode through a capacitance of \(\text{C$_\textrm{inj}$} =\) \SI{242}{\atto\farad} (nominal value).
The amount of injected charge is controlled by an external voltage reference ($V_{\textrm{h}}$) and the injection is triggered via an external signal.

The NMOS source follower output ($OUT2$) is connected to a high-speed (\SI{1.9}{\giga\hertz} unity-gain bandwidth in post layout simulation) 
operational amplifier ($OPAMP$) at the periphery of the matrix that, combined with an output stage capable of driving a \SI{50}{\ohm} load, allows for signal termination on the board or on an oscilloscope to preserve the signal timing information. 
A well-controlled internal feedback loop is used for the operational amplifier to achieve better stability over a wide range of external capacitive loads
\cite{Deng_2023} and the well-controlled output is copied by $M6$ and sent off-chip ($BOARD$) where the external load is.

All voltages and current biases for the front-end and the operational amplifier are externally generated and supplied via pads to the chip, where an ad hoc mirroring ratio is applied to, for example, provide a front-end reset current as low as a few tens of \SI{}{\pico\ampere}.
The in-pixel and out-of-pixel power consumptions per pixel, estimated from the front-end biases, are \SI{150}{\uW} and \SI{3.1}{mW} respectively.

The operating point was optimized using charge injection for the fastest front-end response to establish the best working condition for measuring timing. 
This set of parameters 
is referred to as the operating point, it is given to the chip which distributes it to the pixels. The  operation point values are listed in \ref{apx:voltageSettings}.
Experimental optimization was necessary because the NMOS transistors are in a p-well at voltage \textit{$V_{\rm{sub}}$}, reaching values beyond the range where the transistor model was available. Consequently, an extrapolation of the model was used in simulation to establish the nominal operating point, and proved afterward to deviate significantly from measurements~\cite{Dorda_TTS}.

\subsection{Chip readout system}
\label{sec:chip-readout}
\begin{figure*}[!ht]
    \centering
    \includegraphics[width=\textwidth]{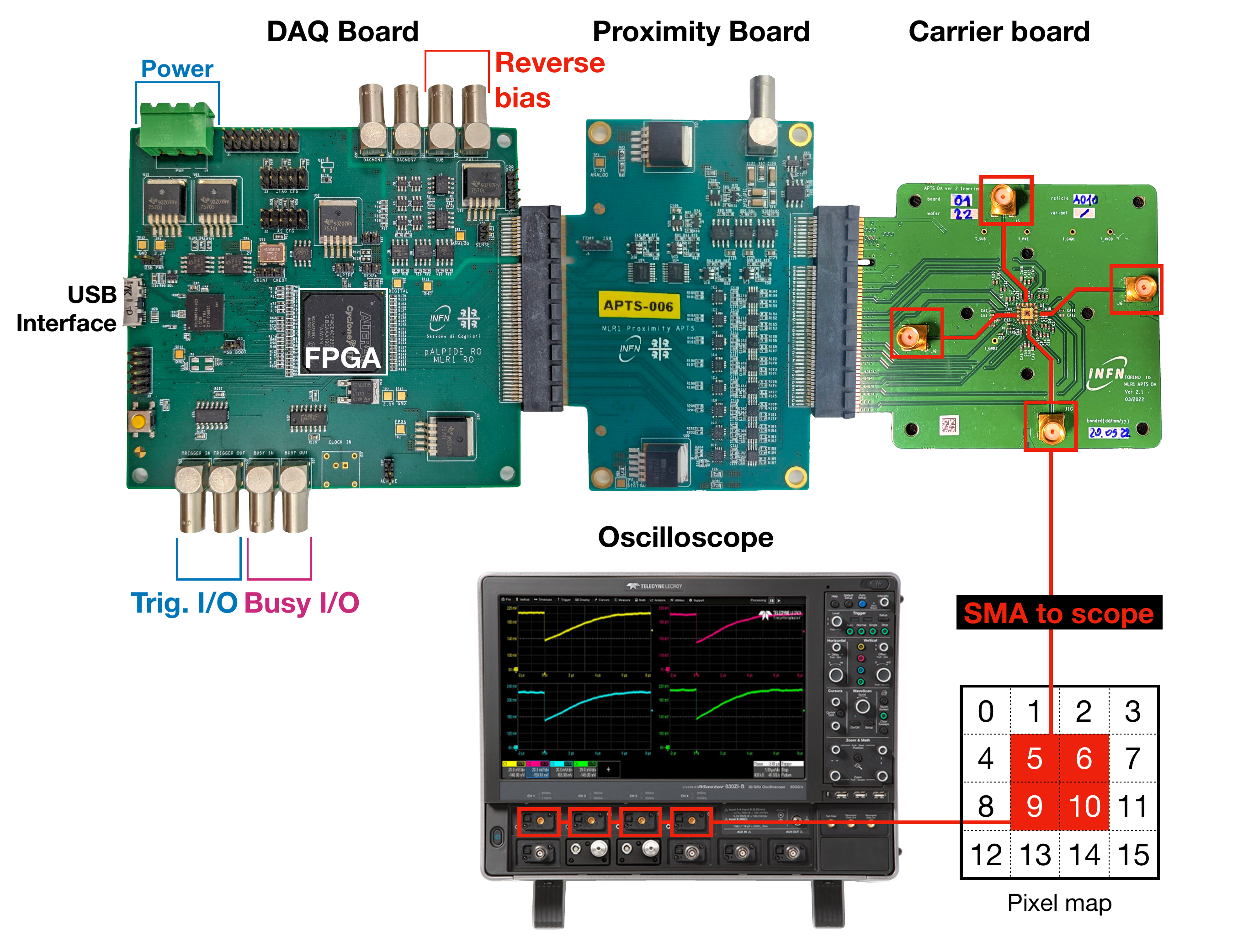}
    \caption{Scheme of the APTS-OA readout system. From left to right: picture of the DAQ board, hosting the firmware control FPGA as well as trigger, busy, and reverse bias connection, the proximity board, hosting ADCs for external pixel readout, and the carrier board to which the chip is glued and wire bonded. In the bottom part, a sketch of the 4$\times$4 pixel matrix with the four inner pixels (highlighted in red) connected to an oscilloscope.}
    \label{fig:labsetup}
\end{figure*}

Figure~\ref{fig:labsetup} shows the test setup used to operate the APTS-OA.
It consists of an FPGA-based data acquisition (DAQ) board, powered at 5 V and controlled by PC via USB interface, a proximity board that hosts the analog to digital converters (ADCs) for the readout of peripheral pixels and sends power and biases to the chip, and a carrier board to which the chip is glued and bonded. More details on this system can be found in Ref.~\cite{Sarritzu_2023}.

The carrier board has four high bandwidth SMA connectors soldered to the output lines of the four innermost pixels (highlighted in red in Fig.~\ref{fig:labsetup}) that can be read out with an oscilloscope (Teledyne Lecroy, WaveMaster 813Zi-B, having a sampling rate of \SI{40}{GS/s} and a bandwidth of \SI{13}{GHz}). 
In the present work, only 3 out of 4 pixels will be read out with the oscilloscope.
The remaining pixels (12 external plus one of the inner) are read out by analog-to-digital converters on the proximity board with a  fixed sampling rate of \SI{4}{MHz}.

\section{Test Beam Measurements} 
\label{sec:timing}
To characterize the performance of the APTS-OA with respect to its time resolution, detection efficiency and spatial resolution, a beam test was conducted at the CERN-SPS H6 facility using positive \(120 \text{ GeV/\it{c}}\) hadrons~\cite{Banerjee:2774716}.

\subsection{Test beam telescope}
\label{sec:TBsetup}
\begin{figure}[!ht]
    \centering
    \input{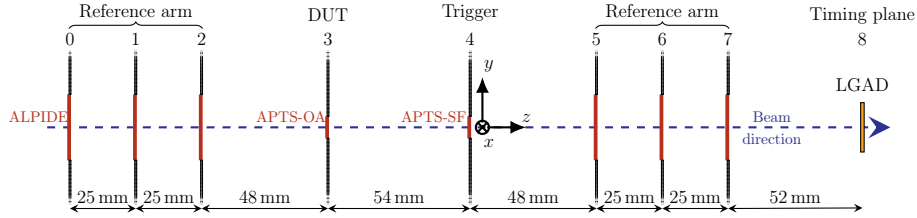}
    \caption{Schematic of the test beam setup showing each detector plane.}
    \label{fig:telescope}
\end{figure}

The setup, which is sketched in Fig.~\ref{fig:telescope}, consisted of nine planes, in particular:
\begin{itemize}
    \item Planes 0, 1, 2, 5, 6, and 7, and 7, each containing an ALPIDE \cite{ALPIDE-proceedings-3},
    tuned to a \SI{100}{e^-} threshold at \SI{-3}{V} reverse substrate bias, for track reconstruction.
    \item Plane 3: an APTS-OA, as a device under test (DUT), placed on xy moving stages.
    \item Plane 4: an APTS-SF \cite{APTSSF_2023}, in the modified process with \SI{15}{\um} pitch is used as trigger plane. 
    Only the central four pixels were activated for the trigger signal generation in order to achieve a coverage slightly larger than the DUT.
    The modified process was chosen to enhance single pixel clusters and therefore the probability to trigger on a seed signal.
    The APTS-SF plane is mounted on xy moving stages and is operated with a \SI{-1.2}{V} reverse substrate bias and all other biases at the optimal operating point, based on Ref.~\cite{APTSSF_2023}.
    \item Plane 8: a Low Gain Avalanche Detector (LGAD) produced by Fondazione Bruno Kessler \cite{Carnesecchi_2023} as a time reference, operated at \SI{110}{V} bias voltage. 
    The signal is amplified with a commercial external circuit (MiniCircuits TB-409-52+) and readout with one channel of the oscilloscope. 
\end{itemize}

The reference system is defined with the positive $z$-axis along the beam direction and with $x$- and $y$-axis defining the sensor planes. 

Two chips having different sensor processes were considered in this study: the modified and the modified with gap variants, as these processes demonstrated the best performance in terms of charge collection \cite{Deng_2023}. 
The substrate bias ($V_{\textrm{sub}}$) was varied from \SI{-1.2}{V} to \SI{-4.8}{V} with steps of \SI{1.2}{V}, consequently varying the depletion and electric field in the sensor. 
Table~\ref{tab:statistic} shows the acquired number of triggers recorded for the different configurations. A larger sample of events was accumulated for the modified with gap structure to study the in-pixel contribution of the lateral electric field.

\begin{table}[!ht]
    \centering
    \begin{tabular}{c|c c }
        $V_{\rm{sub}}$ & Modified with gap & Modified \\
        \hline
        \SI{-1.2}{V} & $290,000$ & $15,000$ \\
        \SI{-2.4}{V} & $334,050$ & $15,000$ \\
        \SI{-3.6}{V} & $15,000$ & $20,000$ \\
        \SI{-4.8}{V} & $258,600$ & $58,100$ \\
        \hline
    \end{tabular}
    \caption{Total number of events recorded during test beam measurements in each configuration.}
    \label{tab:statistic}
\end{table}

\subsection{Data reconstruction and analysis}
\paragraph{DUT signal extraction}

The data recorded in the telescope, which includes the analog waveform from each pixel, was analyzed offline. 
Figure~\ref{fig:signal} shows an example of a typical signal acquired with the oscilloscope. In the inset at the left bottom, the full waveform is shown, while the main plot highlights the main parameters used to extract the amplitude of the waveform:

\begin{figure}[!ht]
    \centering
    \includegraphics[width=\textwidth]{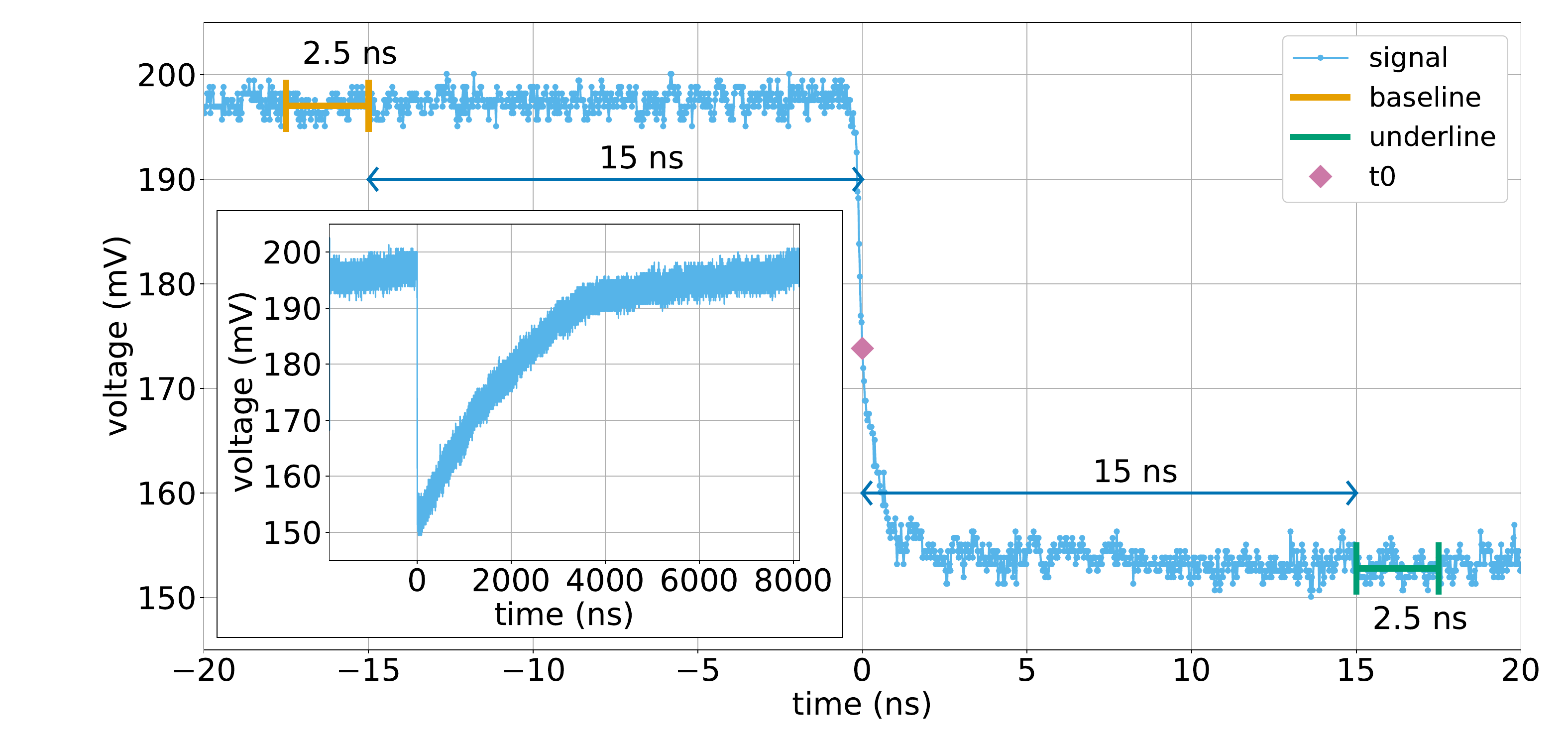}
    \caption{Typical signal acquired by the oscilloscope with baseline, underline, and $t_{0}$ definitions. The $x$-axis is shifted to put $t_{0}$ at the origin to highlight the time intervals for the baseline and underline definitions.}
    \label{fig:signal}
\end{figure}

\begin{itemize}
    \item the time zero (\(t_0\) showed as a pink marker in Fig.~\ref{fig:signal}), estimated as the minimum of the signal's derivative \cite{ugec_2016};
    \item the baseline, calculated as the average of 100 points (corresponding to a time interval of \SI{2.5}{ns}), \SI{15}{ns} before \(t_0\) (orange segment in Fig.~\ref{fig:signal});
    \item the underline, computed as the average of 100 points (\SI{2.5}{ns}), \SI{15}{ns} after \(t_0\), corresponding to the waveform minimum (green segment in Fig.~\ref{fig:signal});
    \item the signal amplitude, determined as the difference between the underline and the baseline;
\end{itemize}
These values are chosen to accurately extract the amplitude of the signal while being robust to noise and the low frequency variation of the baseline.
For the external pixels and one of the 2 $\times$ 2 central ones the signal amplitude is calculated as the difference between the baseline, defined as the 4$^{\rm{th}}$ sample before the minimum of the signal, corresponding to its underline, and the minimum itself, following the definition adopted in Ref.~\cite{APTSSF_2023}.
This definition avoids short-range auto-correlation among points and is not influenced by the signal edge.

\begin{figure}[!ht]
    \centering
    \includegraphics[width=\textwidth]{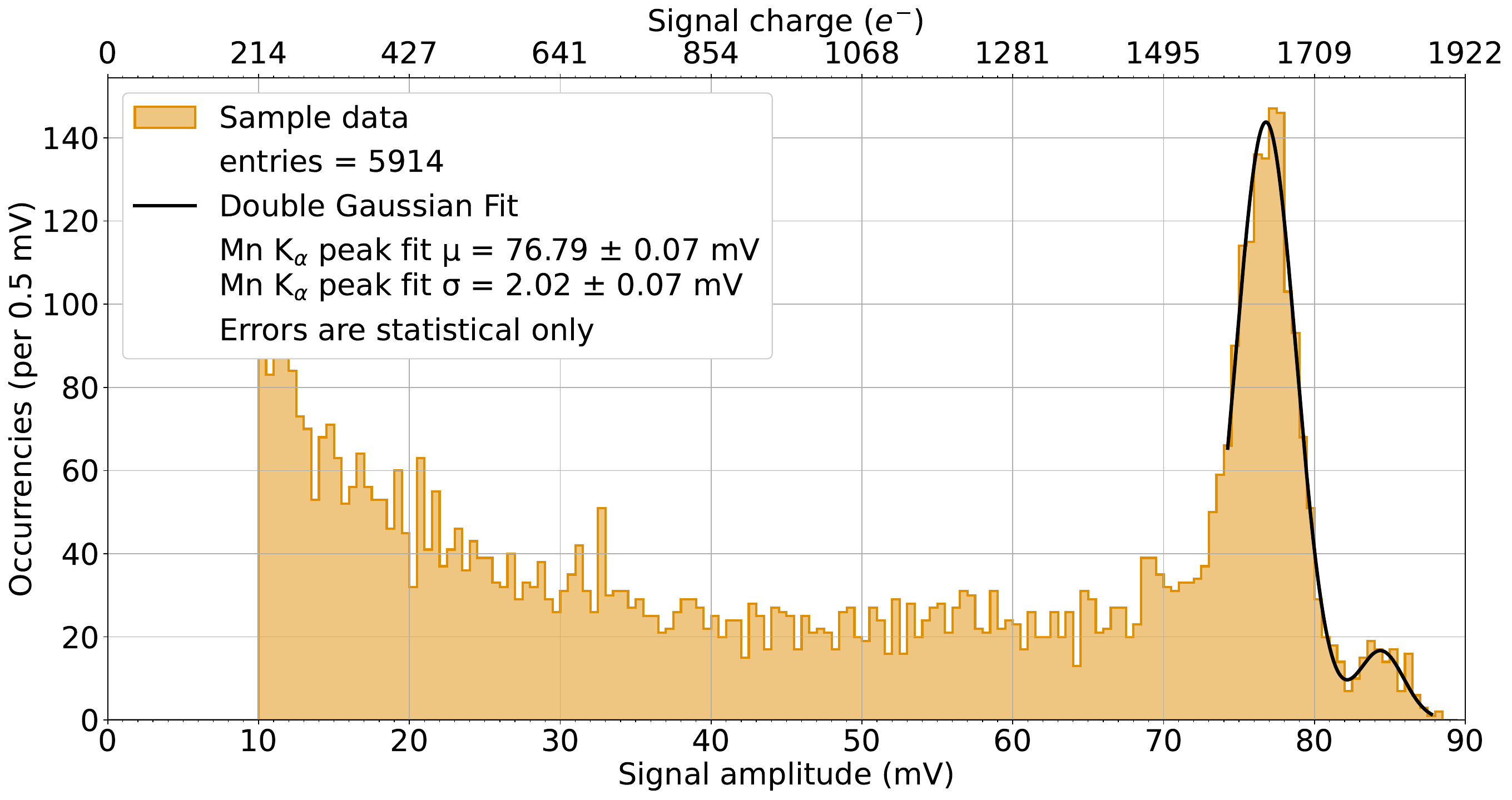}
    \caption{Amplitude distribution of signals acquired with the $^{55}\textrm{Fe}$ source for a single pixel read out with the oscilloscope. The $^{55}\textrm{Mn}$--$K_{\textrm{$\alpha$}}$ and $K_{\textrm{$\beta$}}$ peaks are fitted together with the sum of 2 Gaussian functions.}
    \label{fig:Fe55_spectrum}
\end{figure}

\paragraph{DUT signal calibration}
\label{subsubsec:calibration}
To merge the information from different pixels and compare results taken under different working conditions, the measured signal amplitude is converted into the charge collected by the sensor (expressed in \SI{}{\it{e^-}}). 
The calibration is performed analysing the X-ray spectrum from a $^{55}\textrm{Fe}$ source: the $^{55}\textrm{Mn}-K_{\textrm{$\alpha$}}$ and $^{55}\textrm{Mn}-K_{\textrm{$\beta$}}$ peaks are fitted simultaneously  with 2 Gaussian functions to account for the partial overlap of the peaks at low reverse bias voltages. Then the mean of the $^{55}\textrm{Mn}-K_{\textrm{$\alpha$}}$ peak is used to perform the conversion to electrons with the following formula: 
\begin{equation}
    Q = V_{\rm{out}} \frac{E(K_{\textrm{$\alpha$}})}{V_{\rm{mean}}(K_{\textrm{$\alpha$}}) \epsilon} \ ,
\end{equation}
where $E(K_{\textrm{$\alpha$}})$ is the energy of the $^{55}\textrm{Mn}-K_{\textrm{$\alpha$}}$ photon, $V_{\rm{mean}}(K_{\textrm{$\alpha$}})$ is the mean of the $^{55}\textrm{Mn}-K_{\textrm{$\alpha$}}$ peak estimated from the fit, and $\epsilon$ is the average energy required to produce an electron hole pair in silicon.
For this method the linearity of the energy response of the APTS is assumed, since it has been demonstrated in Ref.~\cite{APTSSF_2023}.\\
The signal amplitude spectrum from a single pixel of the modified with gap sensor operated at $V_{\textrm{sub}}$ = \SI{-4.8}{V} is shown in Fig.~\ref{fig:Fe55_spectrum}. 

\paragraph{Clustering}
Data from DUT, trigger, and tracking planes have been analyzed with the Corryvreckan software framework \cite{kröger2019user}. Adjacent pixels recording a charge above a certain threshold form a cluster. 
The clustering threshold, introduced in the DUT data analysis to emulate the effect of a discriminator, is only applied offline and not during data taking. Each signal amplitude is converted into units of electrons following the calibration procedure described above.
The seed pixel is defined as the one collecting the highest amount of charge within the cluster and the charge-weighted center-of-gravity defines the cluster position.


\paragraph{Track reconstruction}
The Corryvreckan framework uses the General Broken Line (GBL) algorithm~\cite{Blobel:2006yi} to reconstruct the particle tracks by fitting the cluster position on the reference planes and taking into account the multiple scattering effect through the detector material budget. Software alignment of the telescope planes has been performed in subsequent steps in which the different planes are shifted and rotated iteratively with respect to a chosen reference plane to increase the tracking quality.
These tracks are then interpolated at the plane of the DUT.
The six ALPIDE planes and the trigger plane (APTS-SF) are included in the tracking, excluding the DUT in order to obtain unbiased results. 
The following quality selection criteria are applied to filter the data for the analysis:
\begin{itemize}
    \item a cluster is required in each tracking plane;
    \item tracks are selected based on the quality of the fit, such that $\chi^2/{\textrm{dof}} < 5$, where dof stands for the number of degrees of freedom;
    \item only events with a single reconstructed track are selected, where an event has to be triggered by the APTS-SF (cf. Sec.~\ref{sec:TBsetup}, Plane 4);
    \item DUT clusters are associated to tracks passing within a radial distance of \SI{15}{\um}. A study was performed to select the proper association window radius, choosing it in order to guarantee the independence of the efficiency from this parameter (cf. \ref{apx:association_window}).
    
    \item a Region Of Interest (ROI) is set to include only the associated tracks crossing one of the three central pixels read out via the oscilloscope, in order to avoid border effects.
\end{itemize}
The applied selection cuts reduce the statistics to roughly 20\% of the recorded samples quoted in Tab.~\ref{tab:statistic}.

\paragraph{Detection efficiency analysis}
\label{subsubsec:Effic_and_SpatRes} 
In order to estimate the detection efficiency, a cluster is searched on the DUT plane, within the ROI, for each track.
The detection efficiency is computed as the ratio between the number of tracks providing an associated cluster in the DUT and the total number of tracks crossing the ROI. It is worth noting that the choice of the above mentioned association window radius allows one to estimate the efficiency without any bias, taking into account that the tracking resolution is much smaller, being equal to $\sigma_{\textrm{track}}$ = ~\SI{1.8}{\um}~$\pm$~\SI{0.1}{\um}, as obtained by the telescope optimiser tool \cite{mager_telescope}.
Moreover, to assess the chip noise, an analysis of the baseline fluctuations was conducted (cf.~\ref{apx:Noise}). The results in the following 
figures are shown for thresholds above \SI{75}{\rm{e^-}},
3 times the maximum noise RMS.

\paragraph{Spatial resolution analysis}
For this analysis, the so-called $\eta$-algorithm \cite{BUGIEL2021164897,Turchetta:1993vu} has been applied to correct for the non-linear charge-sharing mechanism among the pixels. 
Residual distributions are obtained by evaluating the distance between the intercept of each track on the DUT plane (reconstructed position) and the associated cluster position (corrected with the $\eta$-algorithm), in both the $x$ and $y$ directions. 
The spatial resolutions $\sigma_{x}$ and $\sigma_{y}$ in, respectively, the $x$ and $y$ directions are computed by quadratically subtracting the estimated telescope tracking resolution $\sigma_{\textrm{track}}$ from the standard deviation of the residual distributions in the $x$ and $y$ direction, respectively. 
The final position resolution is then defined as the arithmetic average between $\sigma_{x}$ and $\sigma_{y}$ and it will simply be referred to as spatial resolution in the rest of the paper.

\paragraph{Time resolution analysis} 
\label{subsubsec:time_resolution_analysis}
The sensor time residuals are computed as the difference between the time when the APTS-OA DUT seed pixel signal reaches 10\% of its full amplitude and the time reference provided by the LGAD in the beam telescope (cf. Fig.~\ref{fig:telescope}).
Selecting a constant fraction of the signal amplitude aims at mitigating time walk effects in the time resolution; 10\% proved to provide the lowest RMS of time residual distribution (cf.~\ref{apx:CFD_scan}).
A similar method is applied to the LGAD signals, for which the signal leading edge time at 40\% of its amplitude is chosen as it provides the best time resolution.
Consequently, the distribution of the time residuals, namely the time difference between the signals of the DUT and the time reference, is defined as follows:
\begin{equation} \label{eq:time_residuals}
    \Delta t = t^{\textrm{DUT}}_{10\%} -t^{\textrm{LGAD}}_{40\%} \ .
\end{equation}
Only the seed signals of the three pixels read out by the oscilloscope can be used in the analysis, being the only ones providing precise time information. This further reduces the events previously selected by cluster association via tracking by another 15\%, resulting in a number of available signals for the time resolution analysis equal to 17\% of the number of recorded triggers reported in Tab.~\ref{tab:statistic}.\\
Finally, the LGAD time resolution $\sigma_\textrm{LGAD}=$ \SI{30}{ps}~$\pm$~\SI{3}{ps} \footnote{~The LGAD time resolution was characterized with a positive hadron beam (cf. Ref.~\cite{FCarnesecchi_2023}).} is quadratically subtracted from the standard deviation of the time residuals to calculate the time resolution of the DUT $\sigma_{t}$: 
\begin{equation} \label{eq:time_resolution}
    \sigma_{\textrm{t}} = \sqrt{\sigma^2_{\textrm{$\Delta$\textit{t}}} - \sigma^2_{\textrm{LGAD}}} \ ,
\end{equation}
as the two quantities are assumed to be independent.

\subsection{Results and discussion}
In the following section, the results of the analysis are presented. They are grouped in three main paragraphs and results are presented in the following order: detection efficiency, spatial resolution and cluster size, and, finally, the time resolution.
\paragraph{Detection efficiency}

\begin{figure}[!ht]
    \centering
    \includegraphics[width=\textwidth]{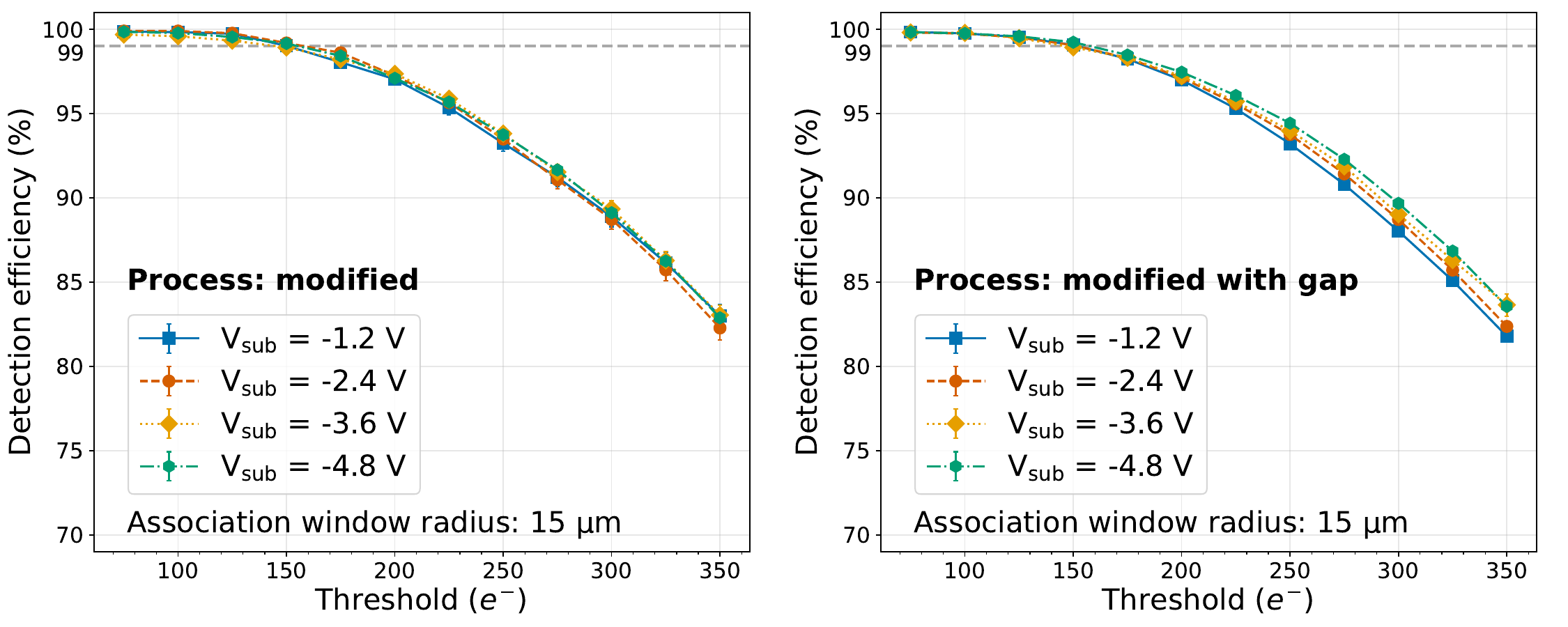}
    \caption{Detection efficiency for the modified (left) and modified with gap (right) pixel processes as a function of the applied threshold and for different V$_{\rm{sub}}$. The dashed line in both plots indicates the 99$\%$ detection efficiency.}
    \label{fig:Eff}
\end{figure}

\begin{figure}[!ht]
    \centering
    \includegraphics[width=\textwidth]{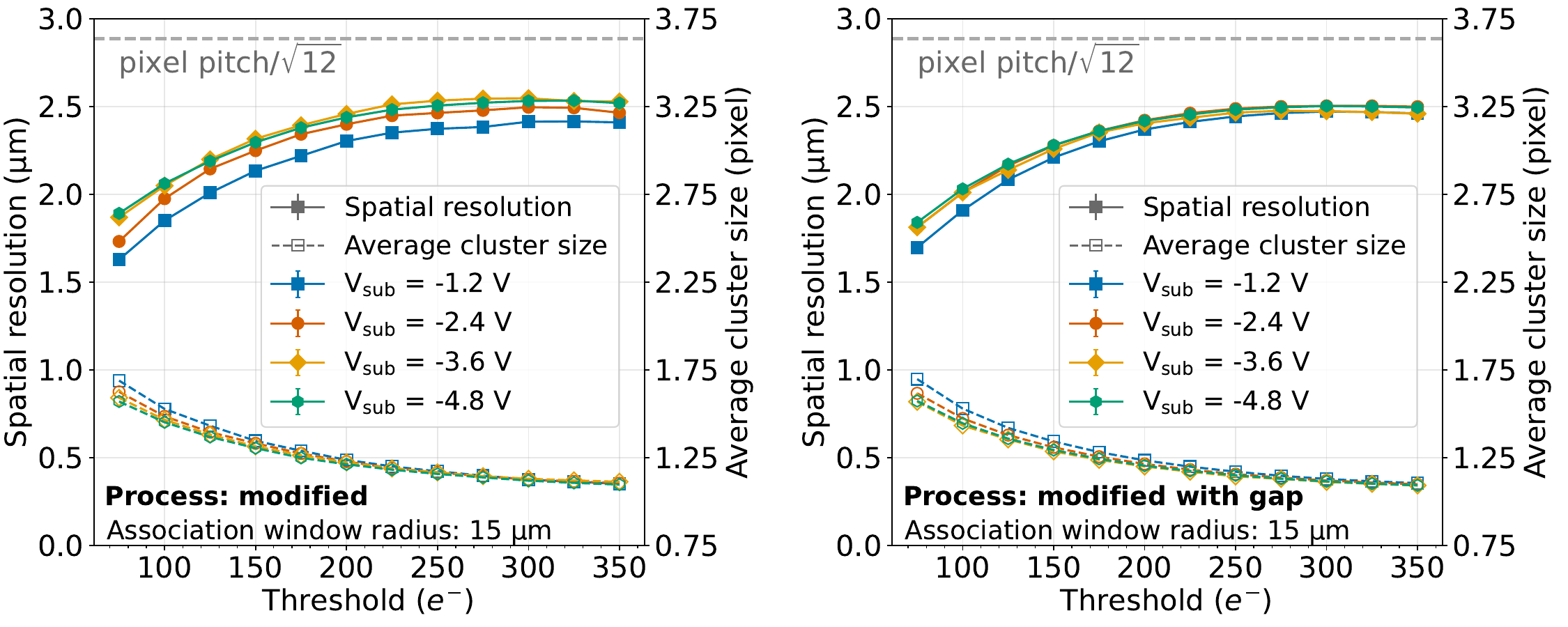}
    \caption{Spatial resolution and cluster size as a function of the applied threshold for the modified (left) and modified with gap (right) structures, comparing different reverse substrate voltages in different colors. The grey dashed line represents the pixel pitch divided by $\sqrt{12}$.}
    \label{fig:spat_res_cs_duts}
\end{figure}

Figure~\ref{fig:Eff} shows the detection efficiency for the modified (left) and modified with gap (right) process as a function of the threshold
at different reverse bias voltages. 
The results are plotted for thresholds exceeding 3 times the RMS of the noise distribution to ensure unbiased efficiency values.
As expected, a clear decreasing trend of the detection efficiency with the increasing threshold can be observed for both the modified and modified with gap process. 


Both pixel variants achieve a detection efficiency higher than 99$\%$ up to a threshold of \SI{150}{\it{e^-}}, in agreement with the results obtained in Ref.~\cite{APTSSF_2023}, offering a comfortable operational range for the sensor. 
Furthermore, achieving a high detection efficiency is essential to prevent any bias in time resolution.
No significant differences are observed between the two variants for this \SI{10}{\um} pixel pitch at the threshold of \SI{100}{\it{e^-}} (cf.~\ref{apx:Efficiency_scan}).


\paragraph{Spatial resolution and cluster size}
Figure~\ref{fig:spat_res_cs_duts} shows the spatial resolution, computed as described in Sec.~\ref{subsubsec:Effic_and_SpatRes}, and the average cluster size as a function of the applied threshold. 
The cluster size decreases with increasing threshold, degrading the spatial resolution, which, nevertheless, for both processes at all biases, is always better than the pixel pitch divided by $\sqrt{12}$.
For both processes, the cluster size also reduces with increasing the reverse bias, because of the enhanced lateral electric field, resulting in a reduced charge sharing. 
This V$_{\textrm{sub}}$ dependence, clearly visible at lower thresholds, gradually becomes undetectable with increasing threshold as the information from low amplitude signals is removed. To better visualize the cluster size, the distributions at different reverse bias are shown in~\ref{apx:CluSiz_distribution}. All the observations are in agreement with the results from APTS-SF (cf. Ref.~\cite{APTSSF_2023}).


\paragraph{Time resolution}

\begin{figure}[!t]
    \centering
    \includegraphics[width=\textwidth]{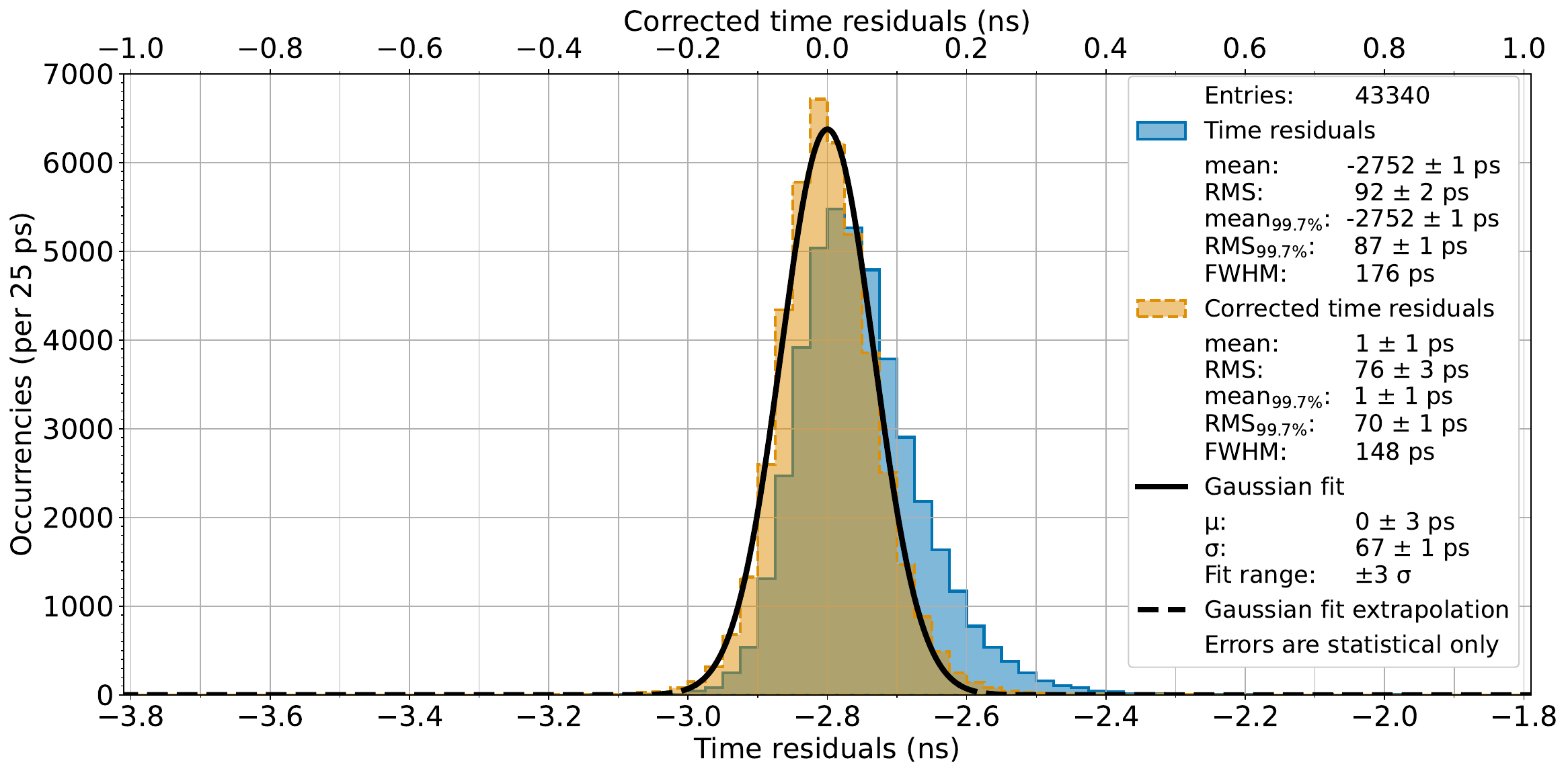}
    \caption{Time residuals measured for the modified with gap process DUT operated at $V_{\textrm{sub}}$=\SI{-4.8}{V}, before (blue distribution, bottom horizontal axis) and after (orange distribution, top horizontal axis) time walk-like correction. The latter distribution is fitted with a Gaussian function on a range of $\pm~3\sigma$ (black solid line, dashed line for points outside the fit range).}
    \label{fig:time_residuals_comparison}
\end{figure}

\begin{figure}[!ht]
    \centering
    \includegraphics[width=\textwidth]{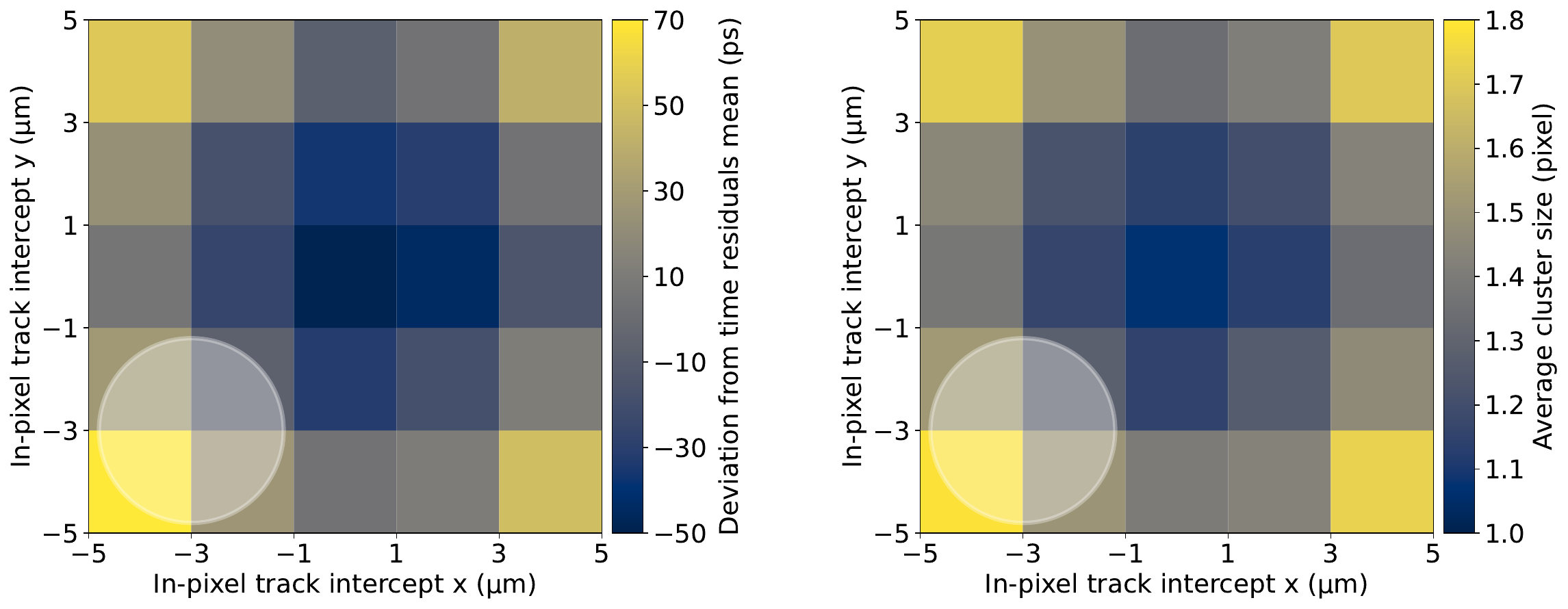}
    \caption{Map of the deviation from the time residuals mean value (left) and cluster size map (right) for in-pixel track incidence positions 
    for the DUT based on modified with gap process operated at V$_{\textrm{sub}}$=\SI{-4.8}{V}. 
    The tracking resolution is represented by the white circle ($r=\sigma_{\textrm{track}}=$ \SI{1.8}{\um}) in the bottom left corner of both maps.}
    \label{fig:inpixel_time_residuals_CS}
\end{figure}

\begin{figure}[!ht]
    \centering
    \includegraphics[width=\textwidth]{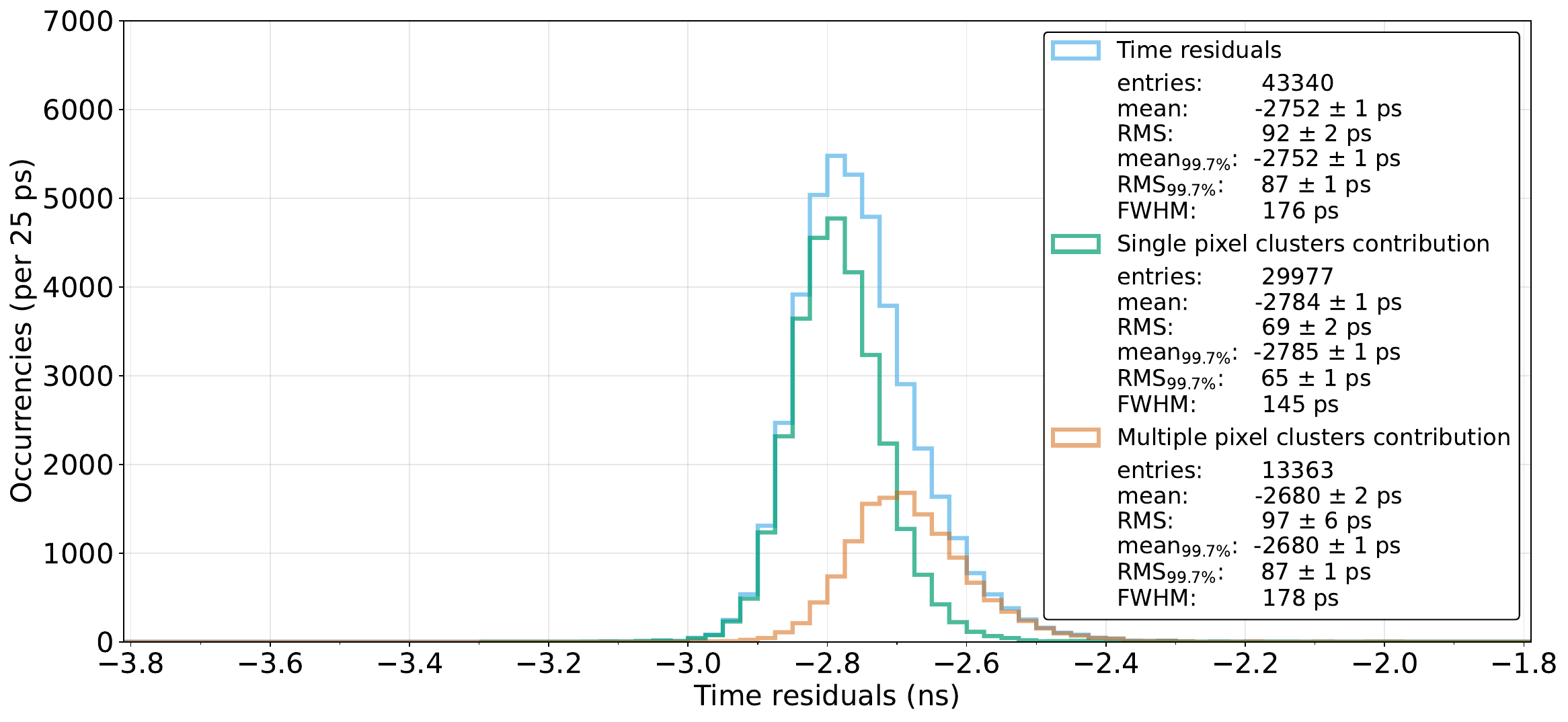}
    \caption{Distributions contributing to non corrected time residuals (blue histogram) divided according to the associated cluster size for the DUT implementing modified with gap process operated at V$_\textrm{sub}$=\SI{-4.8}{V}. The green distribution represents the time residual values related to single pixel clusters, while the red one describes the contribution of clusters with pixel multiplicity larger than one.}
    \label{fig:time_residuals_vs_cluster_size}
\end{figure}

Figure~\ref{fig:time_residuals_comparison} shows, as an example, the time residual distribution for the DUT based on the modified with gap process operated at V$_{\textrm{sub}}$=\SI{-4.8}{V} (blue histogram). The measured time reference signals are delayed with respect to the DUT ones, resulting in negative time residuals (cf. Eq.~\ref{eq:time_residuals}).
\begin{figure}[!ht]
    \centering
    \includegraphics[width=\textwidth]{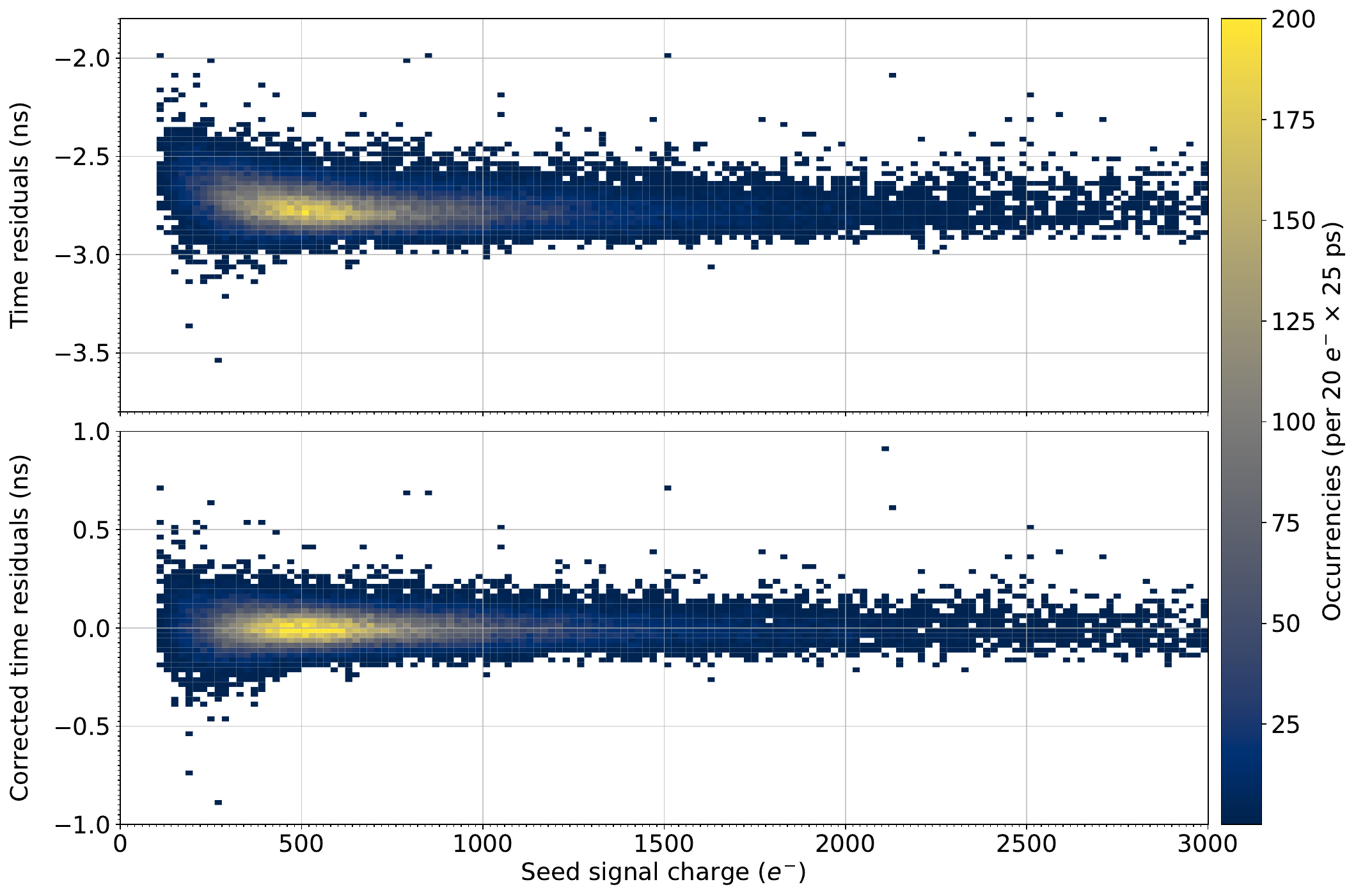}
    \caption{Time residuals versus seed signal charge with no correction applied (top) and after the correction (bottom) for the DUT featuring modified with gap process operated at V$_{\textrm{sub}}$=\SI{-4.8}{V}.}
    \label{fig:pre_post_time_slewing_correction}
\end{figure}\\
Despite the efforts in the analysis to reduce the dependence of the measured signal time on the signal amplitude, the longer tail towards higher time residual values suggests a remaining presence of time walk, confirmed by Fig.~\ref{fig:pre_post_time_slewing_correction} (top) which shows the correlation of time residuals and DUT seed signal charge.
This can be explained by considering that a time measurement based on constant amplitude fraction only suppresses time walk in case of a constant signal shape~\cite{angelo2017}. Since the acquired analog waveforms differ in shape depending on the amount of collected charge and on the position of the ionization within the pixel (cf.~\ref{apx:signal_charge_vs_signal_falltime}), a correction is needed to compensate for these effects.\\
Figure~\ref{fig:inpixel_time_residuals_CS} shows the dependence of the deviation from the mean value of the non-corrected time residual distribution reported in Fig.~\ref{fig:time_residuals_comparison} and of the cluster size on the in-pixel reconstructed track incidence position. 
It can be seen that, on average, lower time residual values, i.e. negative deviations from the mean, are associated to smaller clusters while higher time residuals, i.e. positive deviations from the mean, are associated to larger ones.
In other words, a slower DUT response is observed when a track hits the pixel edges or corners, where the probability of charge sharing with neighboring pixels, hence cluster size, is enhanced. The pixel cluster multiplicity can be therefore used as a proxy to account for the reconstructed track incidence position within the pixel. Single pixel clusters are likely to be related to particles colliding with the center of the pixel, larger clusters with particles hitting the border.
The correlation between time residual values and associated cluster size is further highlighted by Fig.~\ref{fig:time_residuals_vs_cluster_size}. It shows the different contributions coming from signals associated with single pixel clusters 
(green histogram) and from signals associated with multiple pixel clusters 
(red histogram). As already pointed out for Fig.~\ref{fig:inpixel_time_residuals_CS}, the signals associated to single pixel clusters are on average faster and contribute to a narrower distribution than signals from clusters of larger size. Moreover, the distribution associated to the latter explains the asymmetric feature of the time residuals.\\
The arguments presented above suggest a correction to compensate for the dependence of the time residuals on the seed signal charge and on the associated cluster size\footnote{~The performance of the correction based on the dependence of the time residuals on the seed signal charge and on the associated cluster size has been verified with a similar approach based on the dependence on the seed signal charge and on the reconstructed track hit coordinates, with the former producing better results.}, as proposed in Ref.~\cite{Braach_2023}.

For this correction, a clustering threshold of \SI{100}{\it{e^-}} (cf. Sec.~\ref{subsubsec:calibration}) is adopted and the data are divided in two subsets: the first one with single pixel clusters and the second with clusters of multiplicity larger than one.
For both subsets, to avoid introducing any bias in calculating the correction, the mean time residuals for each signal charge bin of one half of the dataset are interpolated with a quadratic function. The latter is then subtracted from the time residuals of the other half of the data set, and vice versa.\\
Figure~\ref{fig:pre_post_time_slewing_correction} (bottom) shows the time residuals as a function of seed pixel signal charge after the correction discussed above. As intended, it reduces the dependence on the seed signal charge and removes its asymmetric feature, as also highlighted by the corrected time residual distribution shown in Fig.~\ref{fig:time_residuals_comparison} (orange histogram).\\
Three estimators for the time residual distribution width are finally derived: the RMS of the entire data set, the RMS evaluated on the central 99.7 percentiles of the distribution and, for the corrected data, the standard deviation $\sigma$ of a Gaussian fit within a 3$\sigma$ interval. The fit is performed on the corrected data only, as, without the correction, the residual distribution is not symmetric around the maximum. The RMS of the corrected distribution gives \SI{76}{ps}~$\pm$~\SI{3}{ps} and is higher than the width $\sigma$=~\SI{67}{ps}~$\pm$~\SI{1}{ps} of the Gaussian fit. The fit is comparable with the RMS$_{99.7\%}$=~\SI{70}{ps}~$\pm$~\SI{1}{ps}. All the reported uncertainties are statistical only.
The time resolution of the APTS-OA is calculated according to Eq.~\ref{eq:time_resolution} by considering RMS$_{99.7\%}$ as the width of the time residual distribution.\\
Since the correction requires a large data sample to be effective, only the data taken with the modified with gap structure operated at V$_{\textrm{sub}}$~=~\qtylist[list-units=single]{-1.2;-2.4;-4.8}{\V} can be corrected.

\begin{figure}[!ht]
     \centering
     \includegraphics[width=\textwidth]{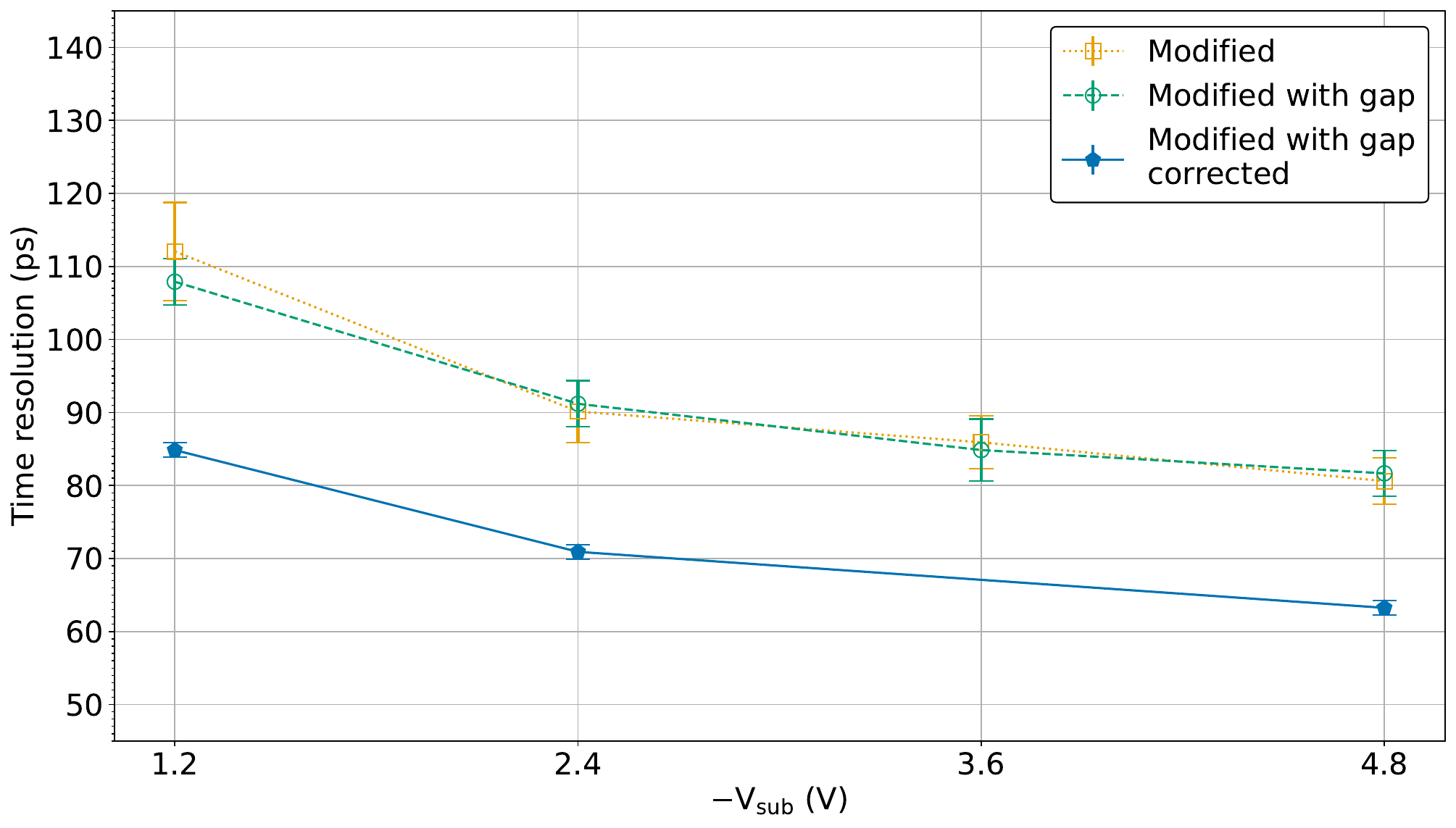}
     \caption{Non-corrected (open symbols, orange dotted line and green dashed line) and corrected (full symbols, blue solid line) time resolution measured for modified and modified with gap DUT process variants. A time walk-like correction, described in Sec.~\ref{subsubsec:time_resolution_analysis}, is applied only to high statistical significance data sets, namely those measured with the modified with gap structure.}
     \label{fig:time_resolution_Vsub_scan}
 \end{figure}

The dependence of the time resolution on the reverse substrate bias voltage V$_{\textrm{sub}}$, for both modified and modified with gap sensors, is shown in Fig.~\ref{fig:time_resolution_Vsub_scan}. 

The time resolution is found to be independent of the process variant,
for this \SI{10}{\um} pitch.
An improvement in timing performance with increasing reverse bias is observed, resulting in a resolution 
$\sigma_{t}$~=~\SI{63}{ps}~$\pm$~\SI{3}{ps} (stat.) and $ \sigma_{t}$~=~\SI{82}{ps}~$\pm$~\SI{3}{ps} (stat.) at V$_{\textrm{sub}}$=\SI{-4.8}{V} for corrected and not corrected data, respectively.

\begin{figure*}[ht!]
    \centering
    \includegraphics[width=\textwidth]{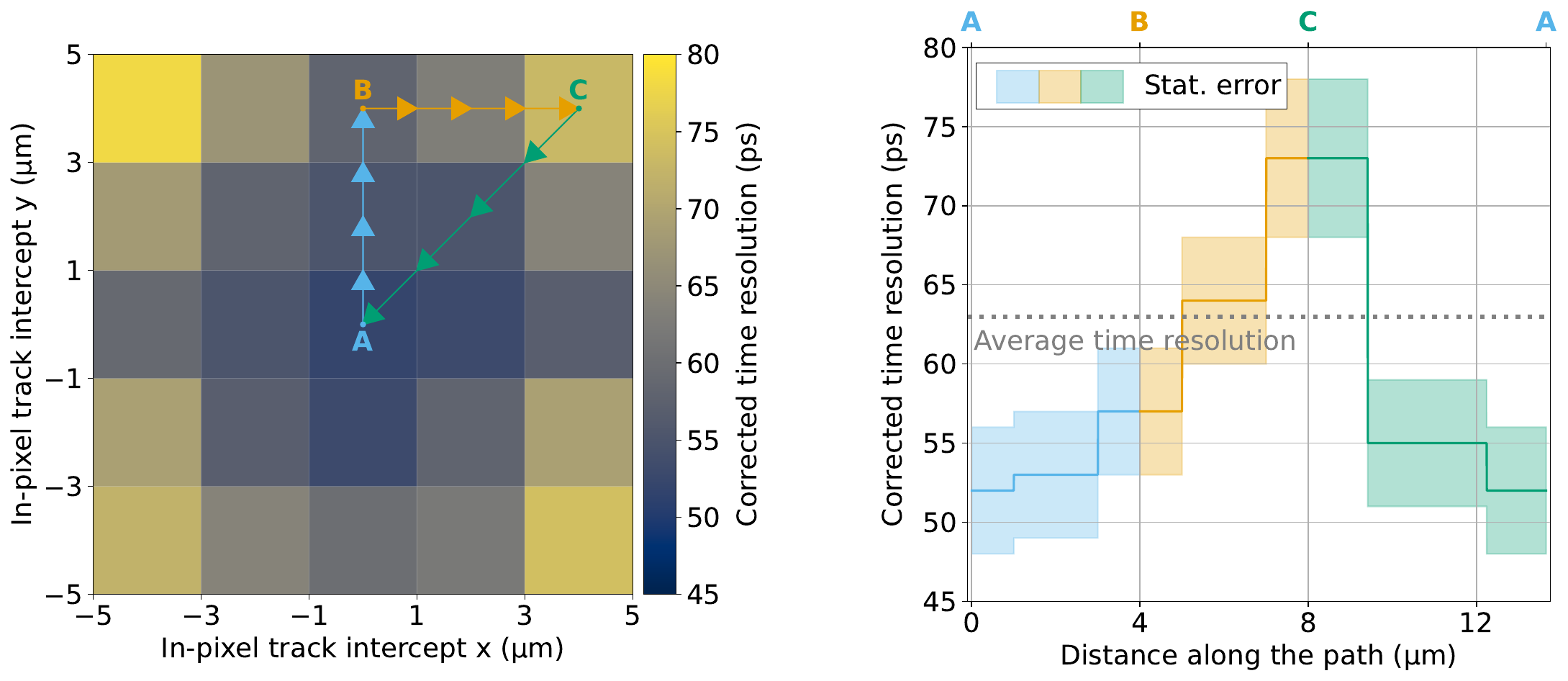}
    \caption{Left: corrected in-pixel time resolution of the modified with gap structure operated at V$_{\textrm{sub}}$=\SI{-4.8}{V}. Right: time resolution variation along the path from the pixel center to the edge, and to the corner.}
    \label{fig:inpixel_time_resolution}
\end{figure*}

The uniformity of timing performance along the pixel surface can be studied by looking at the correlation of the time resolution with the in-pixel track position. Figure~\ref{fig:inpixel_time_resolution} shows the map of corrected time resolution of the modified with gap DUT operated at V$_{\textrm{sub}}$=\SI{-4.8}{V}, with hits from all the pixels grouped per associated track hit position and projected onto one pixel area.
As expected, the lowest time resolution $\sigma_{t}$~=~\SI{52}{ps}~$\pm$~\SI{4}{ps} is measured for tracks hitting the pixel center, where the distance to the collection electrode is minimal and the electric field is maximum, and it deteriorates for incident tracks on the pixel edge ($\sigma_{t}$~=~\SI{58}{ps}~$\pm$~\SI{4}{ps}) and corner ($\sigma_{t}$~=~\SI{72}{ps}~$\pm$~\SI{4}{ps}), marked by the points A, B, and C of Fig.~\ref{fig:inpixel_time_resolution}, respectively. 
The reported uncertainties are statistical only.

The obtained results demonstrate the improvement in particle detection timing performance brought by the TPSCo \SI{65}{nm} CMOS technology with respect to the \SI{180}{nm} CMOS technology. 
In fact, the best time resolution, after the correction, measured for monolithic pixel sensors implemented in \SI{180}{nm} CMOS imaging technology is \SI{106}{ps}~\cite{Braach_2023}, while a time resolution of \SI{63}{ps} is achieved, after time walk correction, for the structures implemented in \SI{65}{nm} CMOS technology discussed here when operated at reverse substrate voltage V$_{\textrm{sub}}$=\SI{-4.8}{V}.\\
However, it is worth noticing that the study presented in this publication concentrates on the time resolution of the sensor only, while the time resolution of a full detector system will also be affected by other components, such as the readout circuitry.

\section{Conclusions}
The performance of the analog pixel test structures realized in the \SI{65}{nm} TPSCo CMOS imaging technology is presented in this publication. 
The APTS-OA was successfully integrated in a beam telescope for the evaluation of its time resolution, spatial resolution, and detection efficiency with minimum ionizing particles.
The modified pixel processes achieved a remarkable time resolution of \SI{63}{ps} with over 99\% detection efficiency and a spatial resolution of \SI{2}{\um} when tested at a substrate voltage V$_{\textrm{sub}}$ of $-4.8\,\textrm{V}$ and \SI{100}{\it{e^-}} threshold, without any observed process dependence.\\
\noindent The results demonstrate the potential of the \SI{65}{nm} CMOS process over its \SI{180}{nm} CMOS process predecessor, especially in terms of sensor development for high-precision time measurements. Monolithic active pixel sensors based on the \SI{65}{nm} CMOS imaging technology are promising candidates not only for future high-energy physics vertex detectors but also for other applications demanding precise timing information in combination with full detection efficiency and high spatial resolution.
\printindex
\newenvironment{acknowledgement}{\relax}{\relax}
\begin{acknowledgement}
\section*{Acknowledgements}
This project has received funding from the European Union’s Horizon Europe research and innovation programme under grant agreement No 101057511.
Funding sources: (1) Project “Dipartimenti di eccellenza” at the Dept of Physics, University of Turin, funded by Italian MIUR.
(2) The National Research Foundation of Korea (NRF) grant funded by the Korean government (MSIT) (No. 2022R1A6A3A03054907)
Project  2022LJT55R, Concession Decree No. 104 of 02.02.2022 adopted by the Italian Ministry of University and Research, cup D53D23002810006.
The measurements leading to these results have been performed at the SPS Test Beam Facility at CERN (Switzerland). We would like to thank the coordinators
at CERN for their valuable support of these test beams measurements and for the excellent test beam environment.
We would like, also, to thank Federico Picollo for providing access to high brillance beam facility for the detector characterization.
ITS3 R\&D and construction is supported by several ITS3 project grants including LM2023040 of the Ministry of Education, Youth, and Sports of the Czech Republic and Suranaree University of Technology, the National Science and Technology Development Agency (JRA-CO-2563-12905-TH, P-2050706), and the NSRF via the Program Management Unit for Human Resources \& Institutional Development, Research and Innovation (PMU-B B47G670091) of Thailand.    
\end{acknowledgement}
%

\bibliographystyle{utphys}
\bibliography{references}

\newpage
\appendix
\section{Analog front-end settings}
\label{apx:voltageSettings}
The operating point used for measurements reported in this paper is summarized in the Table.~\ref{tab:operatingpoints1}.
The currents $I_{\textrm{reset}}$, $I_{\textrm{biasn}}$, $I_{\textrm{biasp}}$ are shown in Fig.~\ref{fig:frontend} while the biasing currents for the operational amplifier $I_{\textrm{bias3}}$, $I_{\textrm{bias4}}$, $V_{\textrm{casn}}$, and $V_{\textrm{casp}}$ are not depicted in Figure. 
Finally, the value of $V_{\textrm{reset}}$ chosen depending on $V_{\textrm{sub}}$ in order to operate in the most linear region of the front-end is reported in Table~\ref{tab:operatingpoints2}.


\begin{table}[!ht]
\centering
\begin{tabular}{ll}
\multicolumn{2}{l}{In-pixel}                      
\\ \hline
\multicolumn{1}{l|}{$I_{\textrm{reset}}$} & $\SI{100}{\pA}$ \\ \hline
\multicolumn{1}{l|}{$I_{\textrm{biasn}}$} & $\SI{-125}{\uA}$ \\ \hline
\multicolumn{1}{l|}{$I_{\textrm{biasp}}$} & $\SI{11.25}{\uA}$ \\ \hline \hline \\

\multicolumn{2}{l}{Out of pixel}                    \\ \hline
\multicolumn{1}{l|}{$I_{\textrm{bias3}}$} & $\SI{212.5}{\uA}$ \\ \hline
\multicolumn{1}{l|}{$I_{\textrm{bias4}}$} & $\SI{2.6}{\mA}$ \\ \hline
\multicolumn{1}{l|}{$V_{\textrm{casn}}$} & $\SI{900}{\mV}$ \\ \hline
\multicolumn{1}{l|}{$V_{\textrm{casp}}$} & $\SI{270}{\mV}$ \\ \hline \hline
\end{tabular}
\caption{The operating point used for the measurement. See the details in Fig.~\ref{fig:frontend}.}
\label{tab:operatingpoints1}
\end{table}

\begin{table}[!ht]
    \centering
    \begin{tabular}{c|c|c|c|c }
    $V_{\textrm{sub}} (\SI{}{\V})$ & $-1.2$ & $-2.4$ & $-3.6$ & $-4.8$ \\
     \hline
    $V_{\textrm{reset}} (\SI{}{\mV})$ & 380 & 420 & 440 & 440
    \end{tabular}
    \caption{$V_{\textrm{reset}}$ settings at different $V_{\textrm{sub}}$}
    \label{tab:operatingpoints2}
\end{table}

\newpage
\section{APTS-OA detection efficiency as a function of the threshold with different association window radius}
\label{apx:association_window}
\begin{figure*}[ht!]
    \centering
    \includegraphics[width=\textwidth]{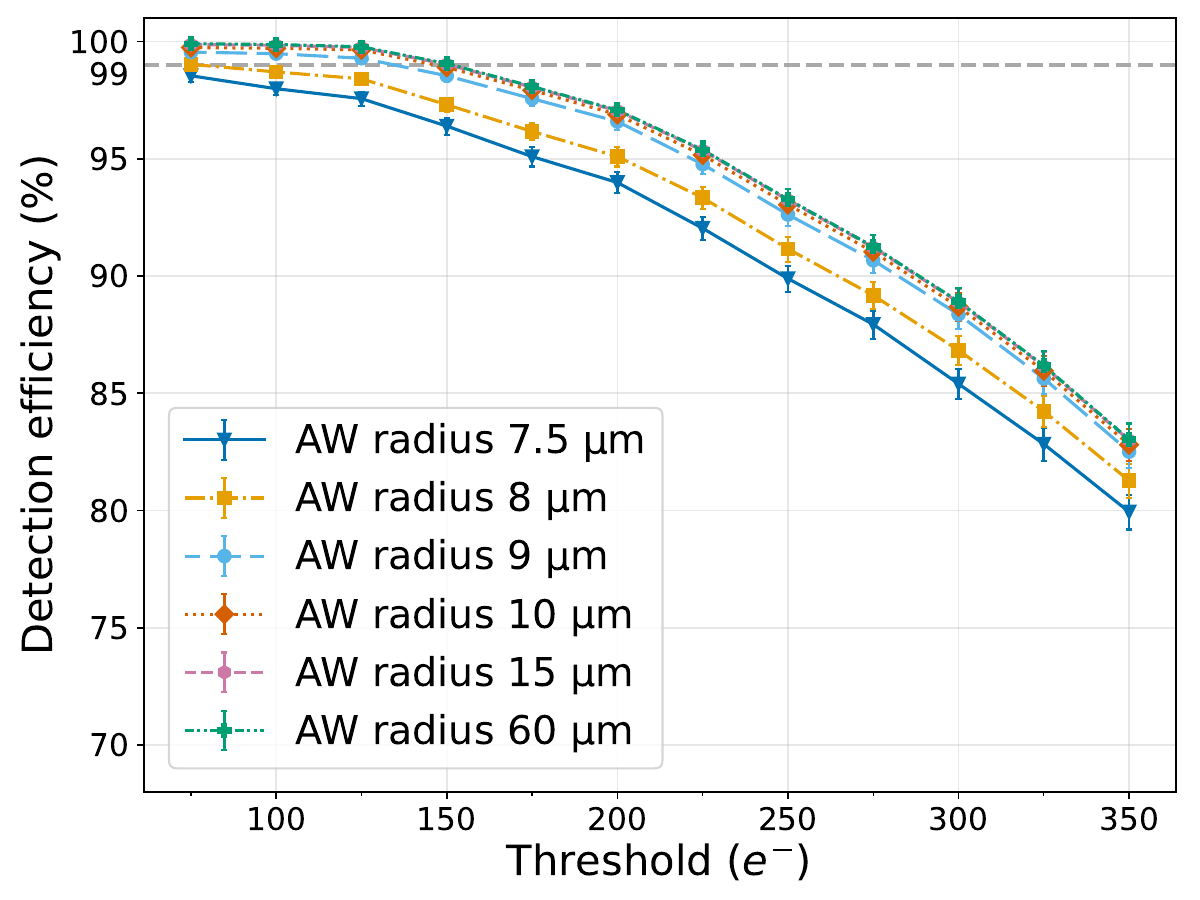}
    \caption{Detection efficiency of the APTS-OA modified DUT as a function of the threshold at V$_{\textrm{sub}}$ = V$_{\textrm{pwell}}$ = \SI{-1.2}{V}, varying the association window (AW) radius. }
    \label{fig:Efficiency_association_window_scan}
\end{figure*}
A study on the association window radius was conducted, starting with a value smaller than the pixel pitch and gradually increasing it to \SI{60}{\um}. The detection efficiency improved with larger association windows across all thresholds. However, the difference in efficiency between the \SI{15}{\um} and \SI{60}{\um} radii was less than 0.05\%. Consequently, an association window of \SI{15}{\um} was selected for the analysis.

\newpage
\section{APTS-OA noise}
\label{apx:Noise}
\begin{figure*}[ht!]
    \centering
    \includegraphics[width=\textwidth]{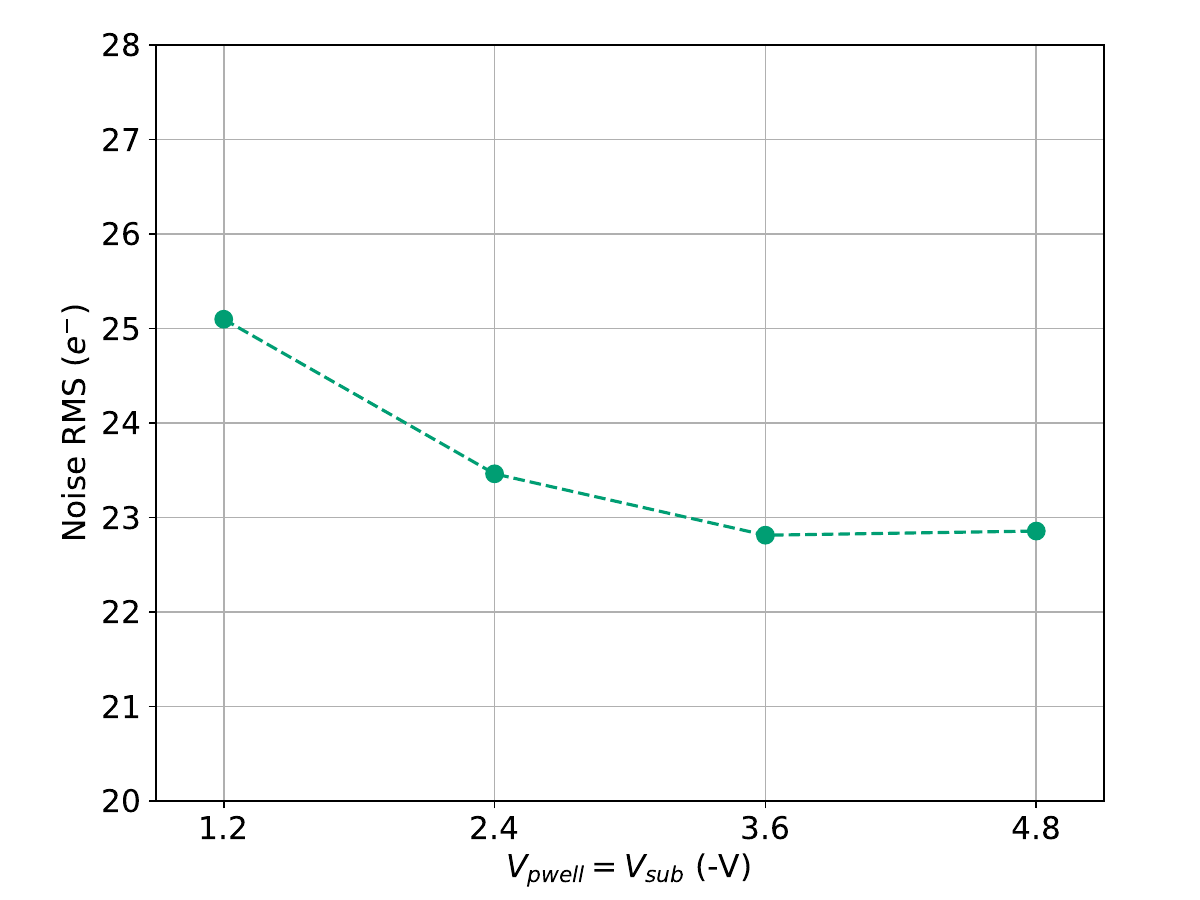}
    \caption{Noise RMS of the APTS-OA modified with gap at different $V_{\textrm{sub}}$ = $V_{\textrm{pwell}}$ configurations.}
    \label{fig:Noise_Vbb_scan}
\end{figure*}
In order to show only results not biased by noise, an analysis of the baseline fluctuations was conducted and efficiency and spatial resolution were plotted for thresholds above 3 times the RMS of the noise
distribution, shown in Fig.~\ref{fig:Noise_Vbb_scan}.

\newpage
\begin{figure*}[ht!]
    \centering
    \includegraphics[width=\textwidth]{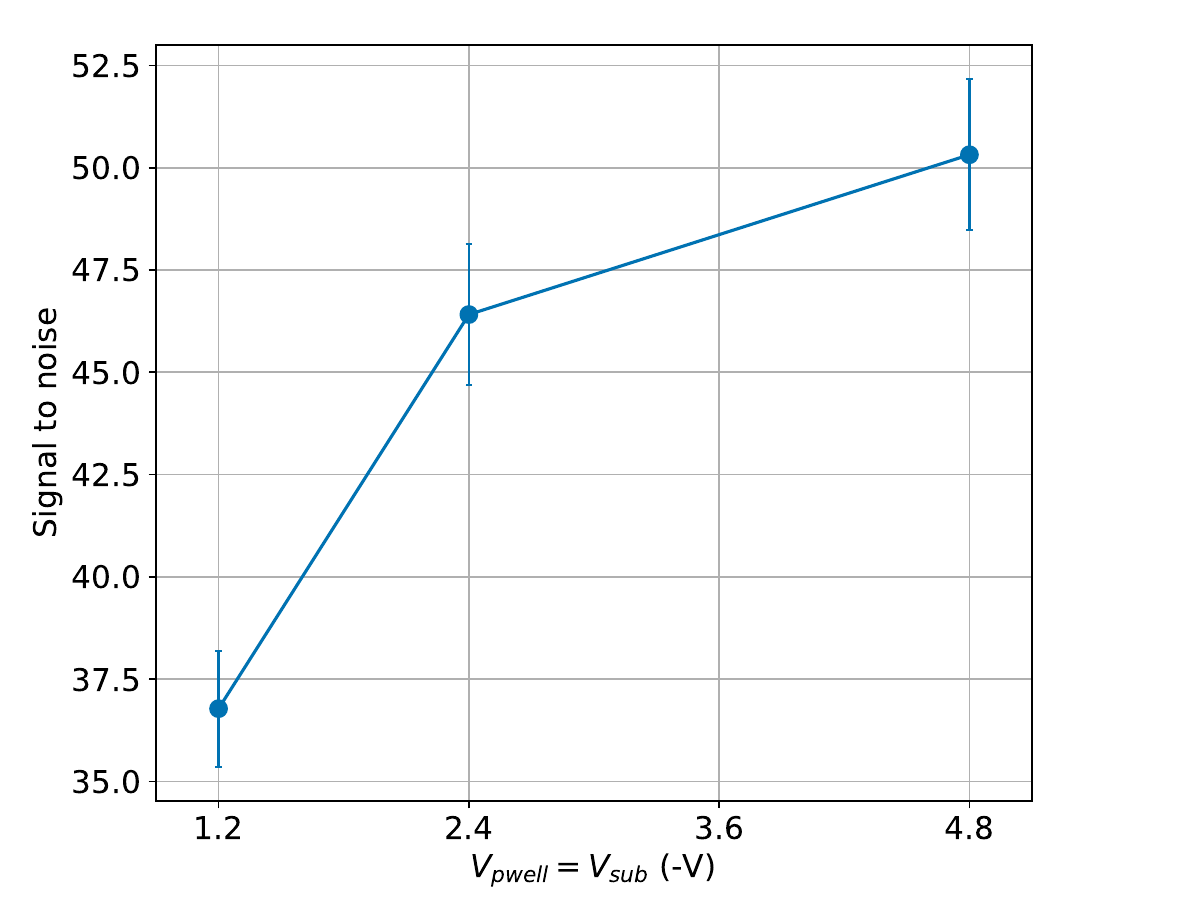}
    \caption{Signal to noise of the APTS-OA modified with gap at different $V_{\textrm{sub}}$ = $V_{\textrm{pwell}}$ configurations.}
    \label{fig:SNR_Vbb_scan}
\end{figure*}
To complete the previous information, the signal to noise ratio (SNR) was calculated for the data samples with the highest statistics ($V_{\textrm{sub}}$ = $V_{\textrm{pwell}}$ = -1.2, -2.4 and -4.8 V) by fitting the matrix charge distribution with a Landau-Gauss function. The most probable value was used to calculate the SNR, shown in Fig.~\ref{fig:SNR_Vbb_scan}.

\newpage
\section{APTS-OA time residuals width at varying signal amplitude fraction}
\label{apx:CFD_scan}
\begin{figure*}[ht!]
    \centering
    \includegraphics[width=\textwidth]{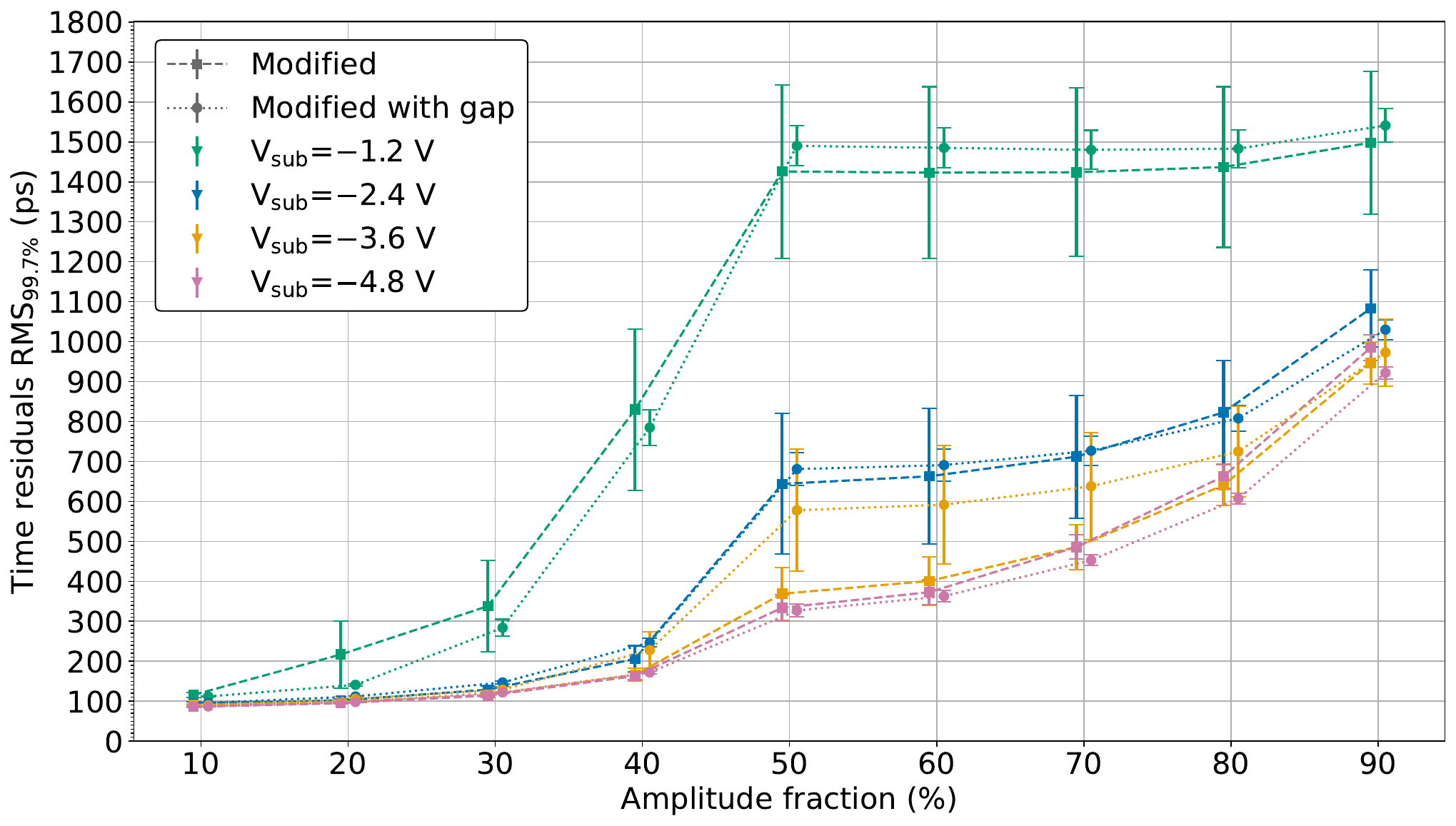}
    \caption{Time residuals distribution width for APTS-OA DUTs' signal times measured at different fractions of the signal amplitude for each operation V$_\textrm{sub}$. The values pertaining to the modified process DUT are indicated with square markers and dashed lines, while those referring to the modified with gap process DUT with circle markers and dotted lines. A slight displacement in the x axis is added between the two curves to allow better visibility.} 
    \label{fig:time_residuals_width_CFD_scan}
\end{figure*}
The time residuals width, expressed as the RMS of the 99.7 central percentiles of the distribution, has been evaluated for both measured DUTs at varying seed signal amplitude fraction. Signal times measured at 10\% of the signal amplitude provide the lowest time residuals RMS, and are therefore used for the analysis presented in this paper.

\newpage
\section{APTS-OA detection efficiency at \SI{100}{\rm{e^-}} threshold}
\label{apx:Efficiency_scan}
\begin{figure*}[ht!]
    \centering
    \includegraphics[width=\textwidth]{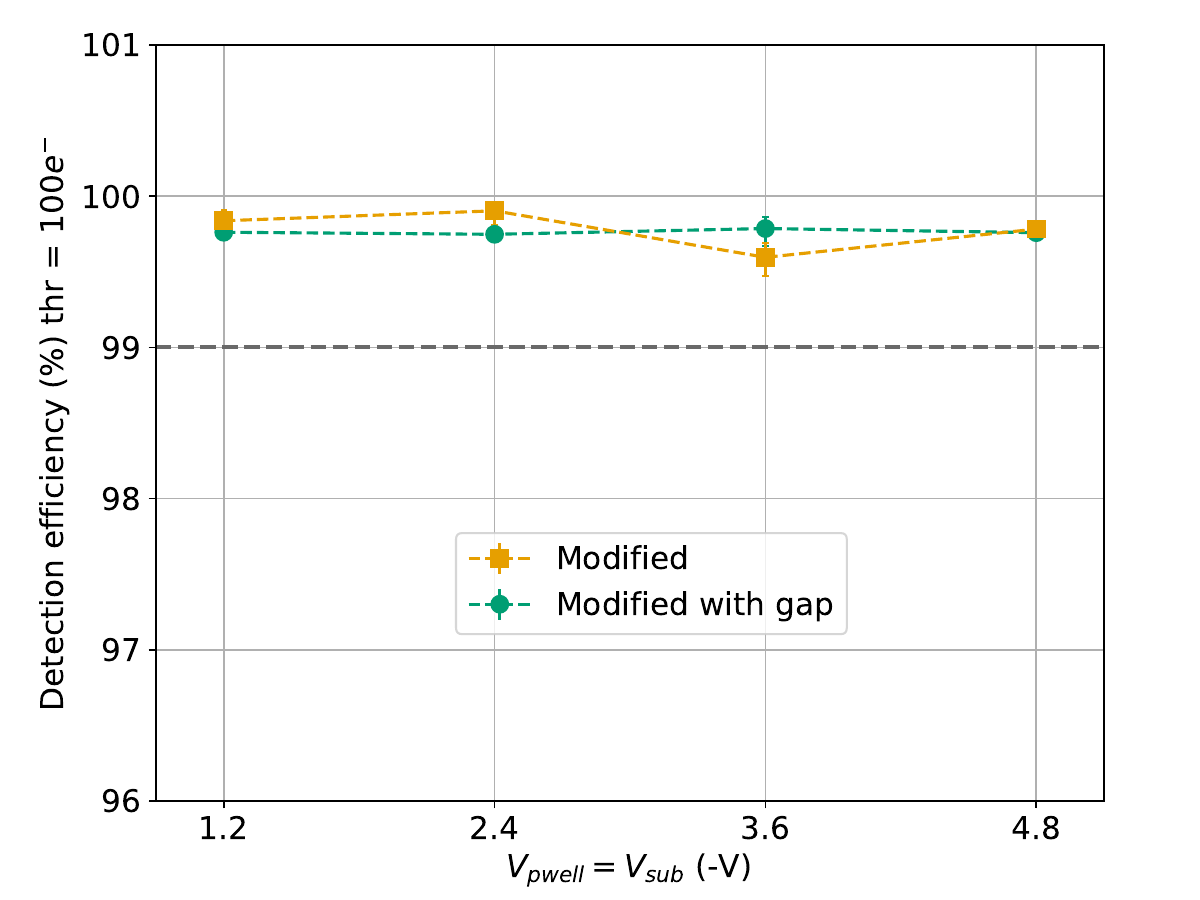}
    \caption{Detection efficiency of the APTS-OA modified and modified with gap DUT at \SI{100}{\rm{e^-}} threshold at different $V_{\textrm{sub}}$ = $V_{\textrm{pwell}}$ configurations.}
    \label{fig:Efficiency_100e_Vbb_scan}
\end{figure*}
Fig.~\ref{fig:Efficiency_100e_Vbb_scan} compares the efficiency measured at \SI{100}{\rm{e^-}} threshold at different $V_{\textrm{sub}}$. No significant difference can be observed between the modified and modified with gap process variants and the efficiency is well above 99$\%$ for all the $V_{\textrm{sub}}$ configurations.

\newpage
\section{APTS-OA cluster size distributions}
\label{apx:CluSiz_distribution}
\begin{figure*}[ht!]
    \centering
    \includegraphics[width=\textwidth]{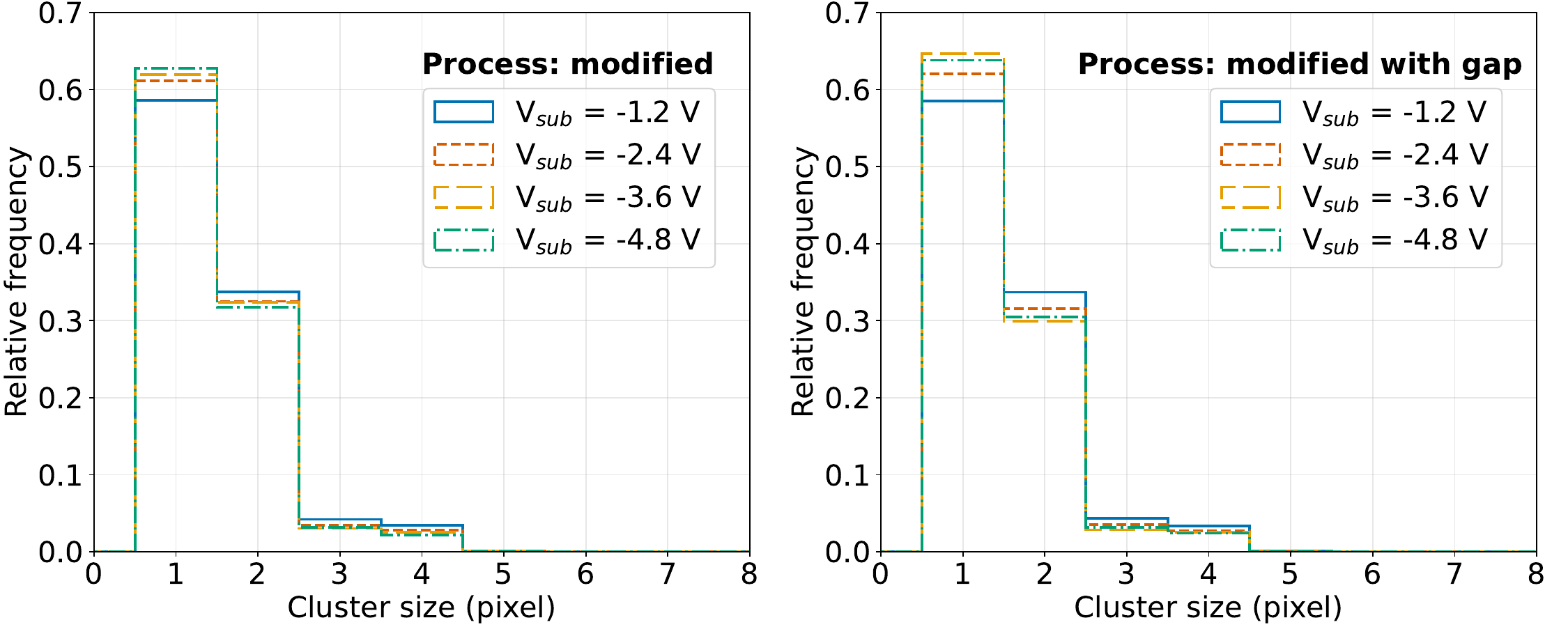}
    \caption{Cluster size distribution of the APTS-OA modified and modified with gap DUT at \SI{100}{\rm{e^-}} threshold at different $V_{\textrm{sub}}$ = $V_{\textrm{pwell}}$ configurations.}
    \label{fig:CluSiz_distribution_100e_Vbb_scan}
\end{figure*}
Fig.~\ref{fig:CluSiz_distribution_100e_Vbb_scan} shows the cluster size distribution at \SI{100}{\rm{e^-}} threshold at different $V_{\textrm{sub}}$ for the APTS-OA modified and modified with gap processes. All the tracks are perpendicular to the DUT plane and the incident angles on x and y are lower than 0.001 degree.

\newpage
\section{APTS-OA correlation between seed signal charge and seed signal leading edge time}
\label{apx:signal_charge_vs_signal_falltime}
\begin{figure*}[ht!]
    \centering
    \includegraphics[width=\textwidth]{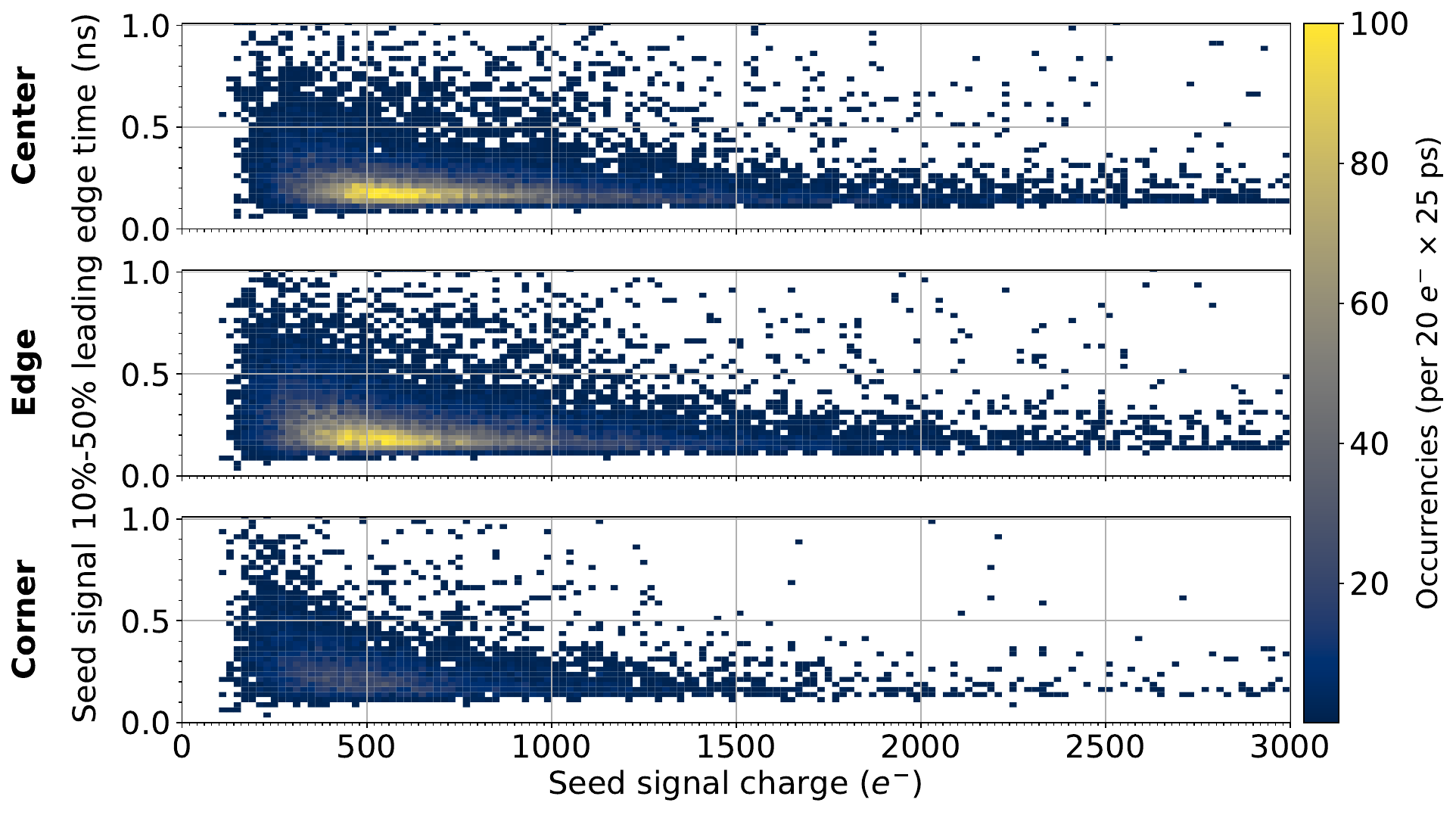}
    \caption{Correlation between seed signal charge and signal leading edge time for the APTS-OA modified with gap DUT operated at V$_{\textrm{sub}}$=\SI{-4.8}{V}. Data is grouped in three different datasets depending on the reconstructed in-pixel track hit possition.}
    \label{fig:signal_charge_vs_signal_falltime}
\end{figure*}
The non-uniformity of the analog signal shape is investigated by studying the correlation between the seed signal charge and the signal leading edge time, expressed as the difference between the signal time at 50\% of full amplitude and the time at 10\% of full amplitude. The leading edge time is used as a proxy to evaluate the signal slope. Slower signals, i.e. longer leading edge times, are observed for smaller collected charge. Data are selected depending on the reconstructed in-pixel associated track hit position. With reference to the \SI{2}{\um}~$\times$~\SI{2}{\um} in-pixel binning used in Fig.~\ref{fig:inpixel_time_residuals_CS}, events associated to the pixel center have reconstructed intercept \SI{-3}{\um}~$<~x~<$~\SI{3}{\um} and \SI{-3}{\um}~$<~y~<$~\SI{3}{\um}. Events associated to the corner dataset belong to the four corner in-pixel bins, edge is the remaining data.

\end{document}